\documentclass[sigconf]{acmart}
\usepackage{subfig}
\usepackage{graphicx}
\usepackage{physics} % for \lVert and \rVert
\AtBeginDocument{%
  \providecommand\BibTeX{{%
    \normalfont B\kern-0.5em{\scshape i\kern-0.25em b}\kern-0.8em\TeX}}}

\begin{document}

%%
%% The "title" 
\title{Grasp Prediction based on Local Finger Motion Dynamics}

%%
%% The "authors" 
\author{Dimitar Valkov}
\affiliation{%
  \institution{Saarland University, DFKI, Saarland Informatics Campus}
  \city{Saarbr{\"u}cken}
  \country{Germany}}
\email{dimitar.valkov@dfki.de}

\author{Pascal Kockwelp}
\affiliation{%
  \institution{University of M{\"u}nster}
  \city{M{\"u}nster}
  \country{Germany}}
\email{pascal.kockwelp@uni-muenster.de}

\author{Florian Daiber}
\affiliation{%
  \institution{DFKI, Saarland Informatics Campus}
  \city{Saarbr{\"u}cken}
  \country{Germany}}
\email{florian.daiber@dfki.de}

\author{Antonio Kr{\"u}ger}
\affiliation{%
  \institution{DFKI, Saarland Informatics Campus}
  \city{Saarbr{\"u}cken}
  \country{Germany}}
\email{krueger@dfki.de}

%%
%% This command allows the author to define a more concise list of authors' names for this purpose.
\renewcommand{\shortauthors}{Valkov et al.}

%%
%% The abstract 
\begin{abstract}
The ability to predict the object the user intends to grasp offers essential contextual information and may help to leverage the effects of point-to-point latency in interactive environments. 
This paper explores the feasibility and accuracy of real-time recognition of uninstrumented objects based on hand kinematics during reach-to-grasp actions.
In a data collection study, we recorded the hand motions of 16 participants while reaching out to grasp and then moving real and synthetic objects. 
Our results demonstrate that even a simple LSTM network can predict the time point at which the user grasps an object with a precision better than 21 ms and the current distance to this object with a precision better than 1 cm. 
The target's size can be determined in advance with an accuracy better than 97\%.
Our results have implications for designing adaptive and fine-grained interactive user interfaces in ubiquitous and mixed-reality environments.
\end{abstract}




\settopmatter{printacmref=false}
\setcopyright{none}
\renewcommand\footnotetextcopyrightpermission[1]{}
\pagestyle{plain}
\maketitle

\section{Introduction}
Hand- and gesture-based natural user interfaces, which allow real or virtual objects to be directly grasped and manipulated, have recently received renewed interest. 
This is by no means surprising since our hands are remarkable instruments that help us shape, transform, use, and manipulate our surroundings. 
However, the widespread adoption of these interfaces is still hindered by issues such as tracking latency and inaccuracy, leading to selection problems, imprecision, and user frustration. 
Hand and finger motions during reach-to-grasp (R2G) have been modeled as a function of the intended object~\cite{Cavallo:2016, robot-grasp-model}, environmental factors~\cite{Santina:2017}, or induced virtual manipulations~\cite{recent-CHI}. 
Such models bear the potential to alleviate the aforementioned problems. 
However, these either describe the hand motion in a scaled, uniform time interval~\cite{recent-CHI,Santina:2017, robot-grasp-model} or require the target object and the intended action to be known in advance~\cite{Clarence:2022, Azmandian:2016:hapticRetargeting, Feix2014grasptask}, which makes them inapplicable for real-time, interactive setups.

One potential approach to enhancing the practicality of these models would be to create an algorithm capable of predicting the target object and estimating the time until the user grasps it. 
For instance, leveraging the known duration of the R2G action alongside the projected remaining time enables the computation of the uniform time parameter essential for the hand redirection model proposed by Gonzalez and Follmer~\cite{recent-CHI}. 
An early forecast of the target object will further enhance the utility of this model, paving the way toward an e.\,g. comprehensive hand redirection technique. 
Furthermore, a precise approximation of the moment when the user contacts an interactive object would enable a proactive initiation of any associated action, effectively minimizing the interaction latency. 
Recent research~\cite{unscr23,unscr25, unscripted, unscr40} has made notable progress in this direction, utilizing the hand motion trajectory. 
However, the unique ability of the hand to convey information about the environment and the user's perception of this environment has rarely been considered~\cite{Vatavu2013a}. 
Indeed, the gradual molding of the fingers during an R2G action has been shown to correlate with the object's physical properties~\cite{Ansuini:2015, Furmanek:2019}, intended use~\cite{Cavallo:2016}, and the user's attitude toward that object~\cite{Ansuini:2006}. 
Unfortunately, while the hand trajectory -- governed by the Fitts' law -- has been a subject of detailed investigations for decades, prehensile kinematics has rarely been considered in the HCI domain.

% in this paper
In this paper, we address this challenge and have collected high-precision data for prehensile hand movements in a controlled data acquisition study. 
Our evaluations indicate that (a) both the current distance to the intended object and the moment at which the user grasps it can be predicted with high precision well before it is reached, and (b) the object can be discriminated during the R2G action if it has sufficiently distinguishable grasping affordances. 
These results have the potential to inform the design of future interactive environments, where virtual objects might be designed in a way that maximizes their discriminability and paves the way toward leveraging the effects of latency by using predictive algorithms able to discriminate future actions.

%%-------------------------------------------------------------------------
\section{Related Work}\label{sc:rw}
Hand prehension is probably one of the most well-studied human activities in the various domains of psychology and neuroscience~\cite{Jones:2006, MacKenzie:1994}. 
Most of the work on hand prehension has focused on investigating the functionality of the hand and the underlying cognitive functions.
For instance, the biomechanical and kinematic properties of the hand and their effect on prehension have been investigated by Chen et al.~\cite{Chen:2011} and Duncan et al.~\cite{Duncan:2013}, some behavioral aspects were addressed e.\,g.\ by Jones~\&~Lederman~\cite{Jones:2006}, and many neuro-psychological models of prehensile behavior have been proposed~\cite{DellaSantina:2017, Santello:2016, Molina-Vilaplana:2002}.
Research groups have also focused on studies on hand pre-shaping, in particular in the R2G task, and have consistently confirmed that grasp formation is highly correlated with the form and the size of the intended object~(see e.\,g.~\cite{Betti:2018} and~\cite{Egmose:2018} for recent surveys), as well as strongly related to the intended action~\cite{Betti:2018, Egmose:2018}. 
Furthermore, Chieffi et al.~\cite{Chieffi:1993} and Ansuini~\cite{Ansuini:2015} investigated the coordination between hand transport and grasp formation and showed that they are mostly independent, but the hand pre-shaping is mostly finalized well before the hand reaches the object~\cite{Chieffi:1993}. 
This observation is also confirmed by the work of Santello~\&~Soechting~\cite{Santello:1998}, Molina-Vilaplana et al.~\cite{Molina-Vilaplana:2002}, and some very preliminary evaluations in the HCI domain~\cite{Daiber2012}. 
Unfortunately, in most cases, the captured motion sequences were trimmed, rescaled to the uniform time interval $[0,1]$, and equidistantly resampled. 
None of these steps can be performed in real time since the entire sequence is required for each of them.
Thus, it is currently not clear how these results can be transferred to practical interactive systems.

% HCI
In the HCI domain, Paulson et al.~\cite{Paulson:2011} have investigated a grasp-shape-based selection of objects in office settings, and Vatavu et al.~\cite{Vatavu2013a} have developed a grasp-posture-based object recognizer for six basic geometric solids that achieved a recognition rate of about 60\% in general, but up to 98\% when using user-specific metrics. 
The correlation between the to-be-grasped object, the task, and the final grasp shape was explored by Feix et al.~\cite{Feix2014graspobject, Feix2014grasptask}.
In contrast, in this paper, we want to explore the possibility of inverting the task--starting with a tracked hand and finger motion, we want to recognize the intended object before the user's hand reaches it.
In this direction, various methods based on hand trajectory have already been proposed. 
These techniques encompass geometric models~\cite{unscr7}, kinematic template matching~\cite{unscr23, unscr40}, regression~\cite{unscr30}, or deep learning~\cite{unscripted, unscr25}. 
Additionally, hand-eye coordination has been explored, leveraging simultaneous hand and gaze tracking~\cite{unscr14, unscr35}. 
When information regarding the intended target is available, target-aware prediction approaches, as suggested by Ahmad et al.~\cite{unscr3} or Yu et al.~\cite{unscr48}, can be utilized. 
Despite the differences among these techniques, they all share the common characteristic of relying solely on the hand movement trajectory, typically in a global tracking space, without leveraging the valuable information provided by the intricate finger motions involved in grasping.

%tracking
The high speed and dexterity of the hand make its precise tracking a challenging task for any tracking approach. 
Researchers have addressed the problem by developing complex ML~\cite{Cheng:2016}, optimization~\cite{Taylor:2016}, or rule-based~\cite{Hoell:2018} algorithms for hand pose reconstruction and gesture recognition~\cite{Cheng:2016}. 
This may lead to misinterpretations and data leakage when used in hand kinematics studies~\cite{Fu:2011}. 
Similar to Heumer et al.~\cite{Heumer:2008}, we based our evaluation on geometric properties extracted directly from the captured motion data. 
This has the additional benefit that similar features can be extracted with alternative, lightweight approaches, such as the Finexus~\cite{Chen:2016}, CyclopsRing~\cite{Chan:2015}, or FingerPad~\cite{Chan:2013}.
Finexus~\cite{Chen:2016}, in particular, demonstrates how the position of the fingertips can be tracked in 3D space with high precision and speed, using only low-cost hardware.

% --------------------------------------------------------------------------------------
\section{Prediction Model}\label{sec:analysis-methods}
\begin{figure}[!b]
    \centering
	  \includegraphics[width=0.98\linewidth]{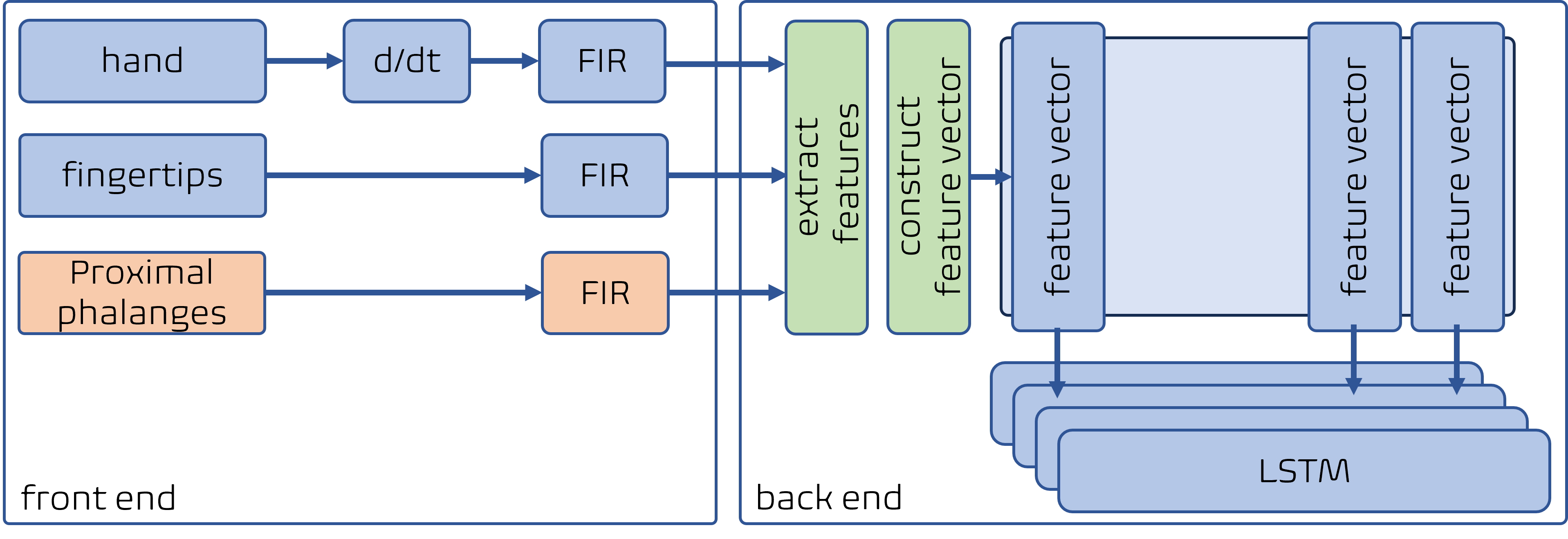}
	\caption{Illustration of the prediction model. The raw sensor data is filtered with FIR filters and used to construct feature vectors. These are then combined in fixed-length sequences and fed to the set of LSTM-NNs for prediction.}
    \Description{}
    \label{fig:overall-model}
\end{figure}
Our prediction model, illustrated in Figure \ref{fig:overall-model}, is built upon a Long Short-Term Memory Neural Network (LSTM-NN) and consists of a hardware-dependent front end and a prediction back end. 

In the front end, the system consistently gathers hand-tracking samples. Here, different methods -- from high-end optical or electromagnetic tracking systems, through computer vision algorithms, through data gloves, to dedicated lightweight hardware solutions, such as Finexus~\cite{Chen:2016} or CyclopsRing~\cite{Chan:2015} -- can be used. The critical aspect is that the front end must provide at least the hand speed, orientation, and the relative position of the fingertips to the remainder of the system. Optionally (s. Section~\ref{sec:eval}), the positions of the proximal phalanges or the proximal interphalangeal joints can also be tracked and provided to the backend.

In the back end, the data is used to extract feature vectors from each tracking frame (see Section~\ref{sec:features}).
These feature vectors are then combined in short sequences of fixed length using a FIFO queue and fed to the LSTM networks. 
These networks are used to predict \emph{the distance} between the hand and the to-be-grasped object, \emph{the time} until the grasp occurs, and the \emph{object's shape or size}. 
Once a sequence of sufficient length is obtained, the prediction can be initiated. 
After that, new predictions can be triggered for each incoming sample, facilitating real-time forecasting.

% Hardware and setup
\subsection{Hardware Setup}\label{sec:hardware}
For the purpose of this work, we wanted to avoid any hardware-related confounding factors and concentrate on the opportunity to extract meaningful information from the hand and finger motion dynamics. 
Therefore, we used at the front end a Polhemus Viper16 electromagnetic tracking system with $12$ tethered micro sensors to track participants' hands and fingers with submillimeter precision at $960$ fps. 
We attached $5$ sensors to the fingernails and $5$ sensors in the middle of the proximal phalanges of the fingers (cf. Figures~\ref{fig:model} and \ref{fig:setup}). 
Two additional sensors were attached to the metacarpal of the thumb and to the metacarpal of the middle finger, respectively. 
The data capturing application was implemented in C/C++ using the Qt framework and was run on a Windows 11 laptop with a Core i7-11800H processor, $32$ GB RAM, and Nvidia RTX 3080 GPU. 

% preprocessing and filtering
\subsection{Preprocessing and Filtering}\label{sec:preprocessing}
The absolute hand movement velocity $\nu_h(t)$ was calculated using the finite difference approximation:
\begin{align*}
    \nu_h(t) = f\cdot\left\lVert p_h(t_k)-p_h(t_{k-1})\right\rVert
\end{align*}
Here, $p_h(t_k)$ denotes the hand position in frame $k$, and $f$ is the tracker's frame rate ($f=960$ Hz in our setup). 
Missing frames are a frequent event in most high-speed trackers. 
While almost invisible in the positional recordings, they manifest as distinctive spikes in the velocity profiles. 
However, these can be easily detected in real-time by examining the local acceleration profile. Specifically, if the accelerations of the $k$-th sample $a_h(t_{k})$
is \emph{above} a certain positive threshold ($0.1$ in our case) \emph{and} the subsequent acceleration $a_h(t_{k+1})$
is \emph{below} a negative threshold ($-0.1$ in our case) then the velocity $v_h(t_k)$ was replaced by the local mean value.

Filtering of tracked positions is an often overlooked problem, with many researchers commonly employing zero-phase, Infinite Impulse Response (IIR) low-pass filters. 
Indeed, zero-phase filtering is problematic as it is an anti-causal process that cannot be implemented in real-time applications. 
Additionally, while IIR filters offer effective noise cancellation, they introduce undesirable phase distortions because of their non-linear, frequency-dependent group delay. 
On the other hand, directly using the unfiltered input carries the risk of the machine learning algorithms being biased by the sensor noise, especially when working with small training sets.

For our setup, we utilized a set of 25th-order, Type 1 Hamming window FIR filters~\cite{elliott1987handbook} with a cut-off frequency of $25$ Hz, due to their stability and strictly linear phase characteristics guaranteeing a constant group delay. 

% feature extraction
\subsection{Feature Extraction}\label{sec:features}
\begin{figure}[!t]
    \centering
	  \includegraphics[width=0.75\linewidth]{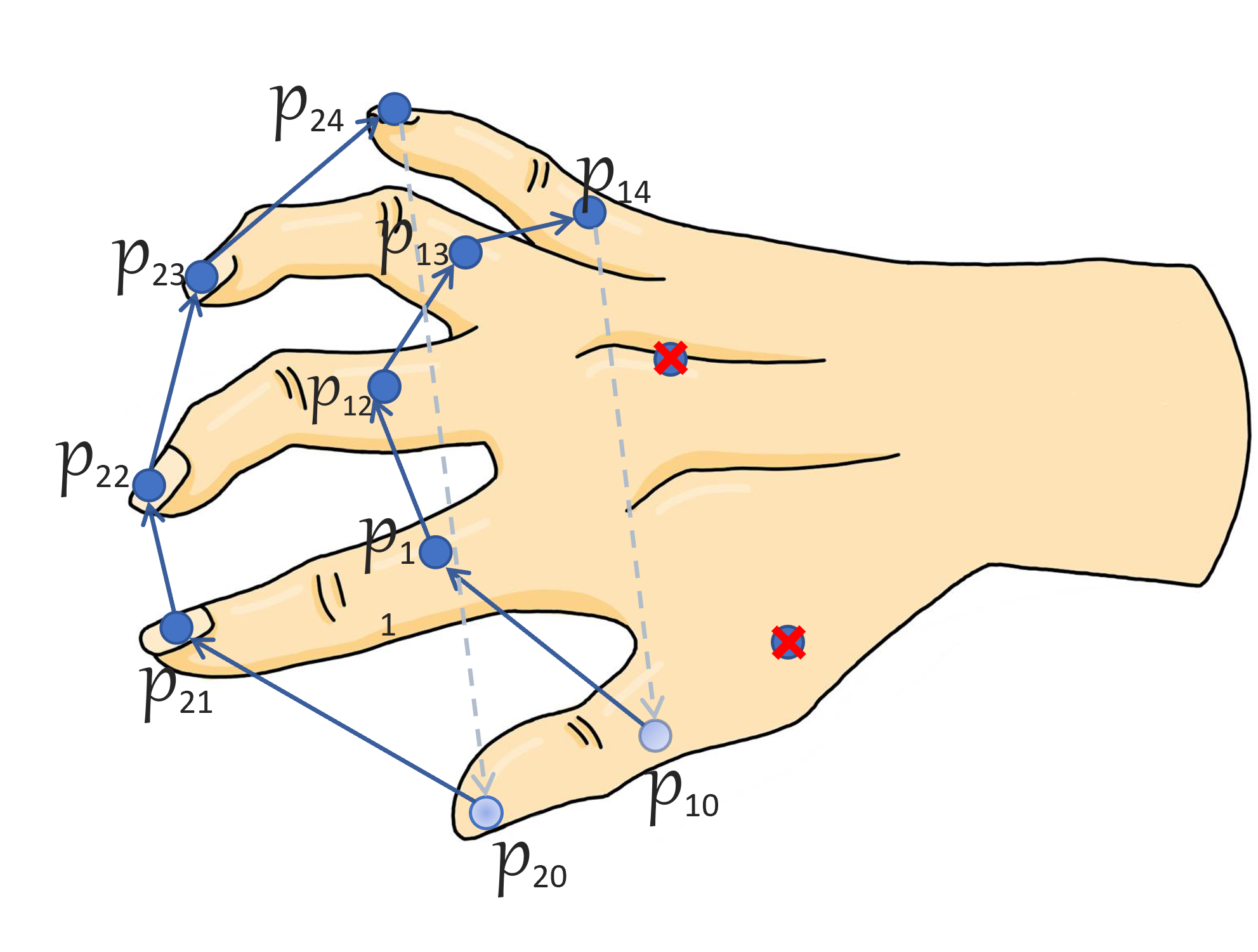}
	\captionof{figure}{Illustration of the hand polygon model. The finger polygon vectors are calculated from the raw sensor positions $p_{ij}$ and used together with the hand motion speed $\nu_h$ to construct a feature sequence.}
    \Description{}
    \label{fig:model}
\end{figure}
The thumb-to-index grip aperture is commonly employed to describe a grasp pose~\cite{Castiello:2005:review}. 
An alternative is to utilize the surface, extend, or roundness of the (non-planar) 3D polygon formed by the fingertips~\cite{Sing:1994}. 
Drawing inspiration from these methods, we formulate the \emph{hand polygon model} (see Figure~\ref{fig:model}), which represents the grip with two independent polygons. 
The FP polygon is constructed using sensors attached to the user's fingertips, while the PP polygon has its edges in sensors attached to the proximal phalanges. 

Our hand polygon model incorporates all important grasp shape features - the thumb-index and thumb-little finger apertures, the relative curvature of the grip, the hand orientation, etc., while removing the global hand position. Notably, all these features strongly correlate with the intended object while being relatively independent of the user's hand and finger size. For instance, regardless of whether the user's fingers are large or small, the thumb-index aperture will be slightly larger than the size of the to-be-grasped object at the end of the R2G motion.

Additionally, we incorporated the hand movement velocity $\nu_h$ into our feature vector. 
The feature vectors are then combined in short temporal sequences with a fixed number of samples, as illustrated in Figure~\ref{fig:overall-model}.

\subsection{Network Architecture}\label{sec:network}
\begin{figure}[!b]
    \centering
    \includegraphics[width=0.95\linewidth]{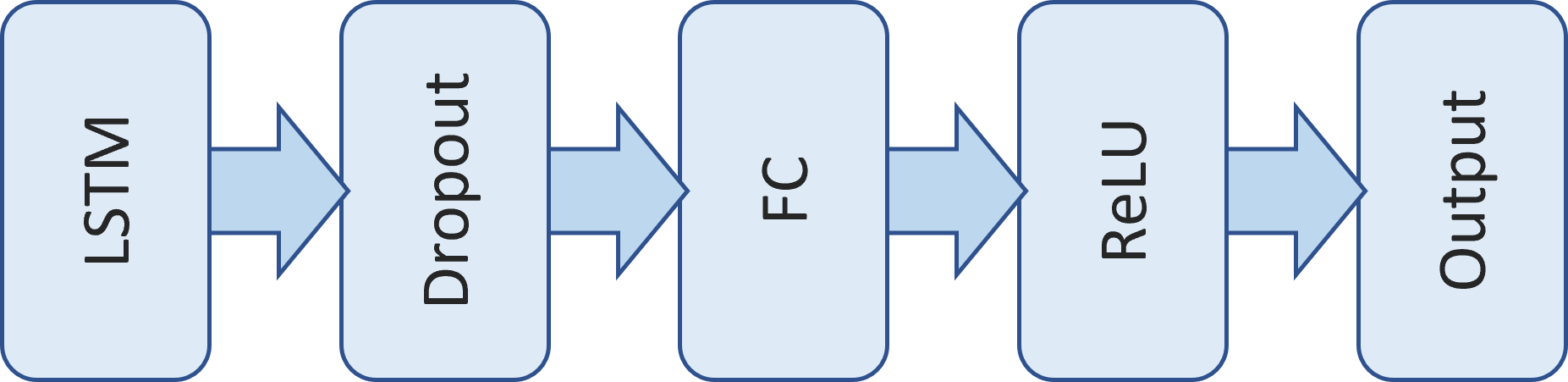}
	\captionof{figure}{Architecture of the LSTM-NN used in our evaluations. The output layer is a single neuron with linear activation for the regression analysis or an FC layer with \emph{softmax} activation for the classification tasks.}
    \Description{}
    \label{fig:nn}
\end{figure}
\begin{figure*}[!ht]
	\centering
	\subfloat[][]{
	    \includegraphics[width=0.23\linewidth]{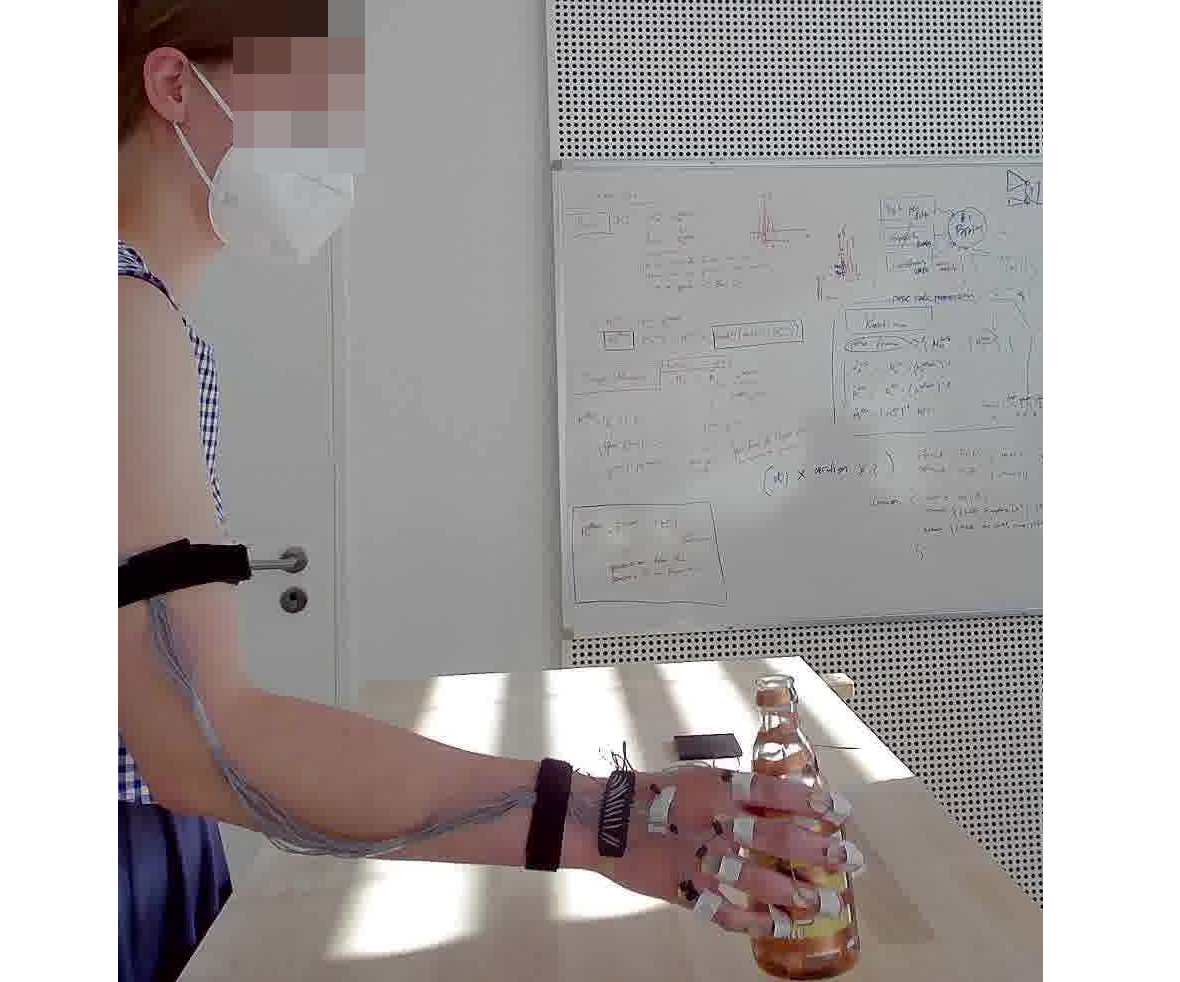}
		\label{fig:participant}
	}
	\subfloat[][]{
	    \includegraphics[width=0.23\linewidth]{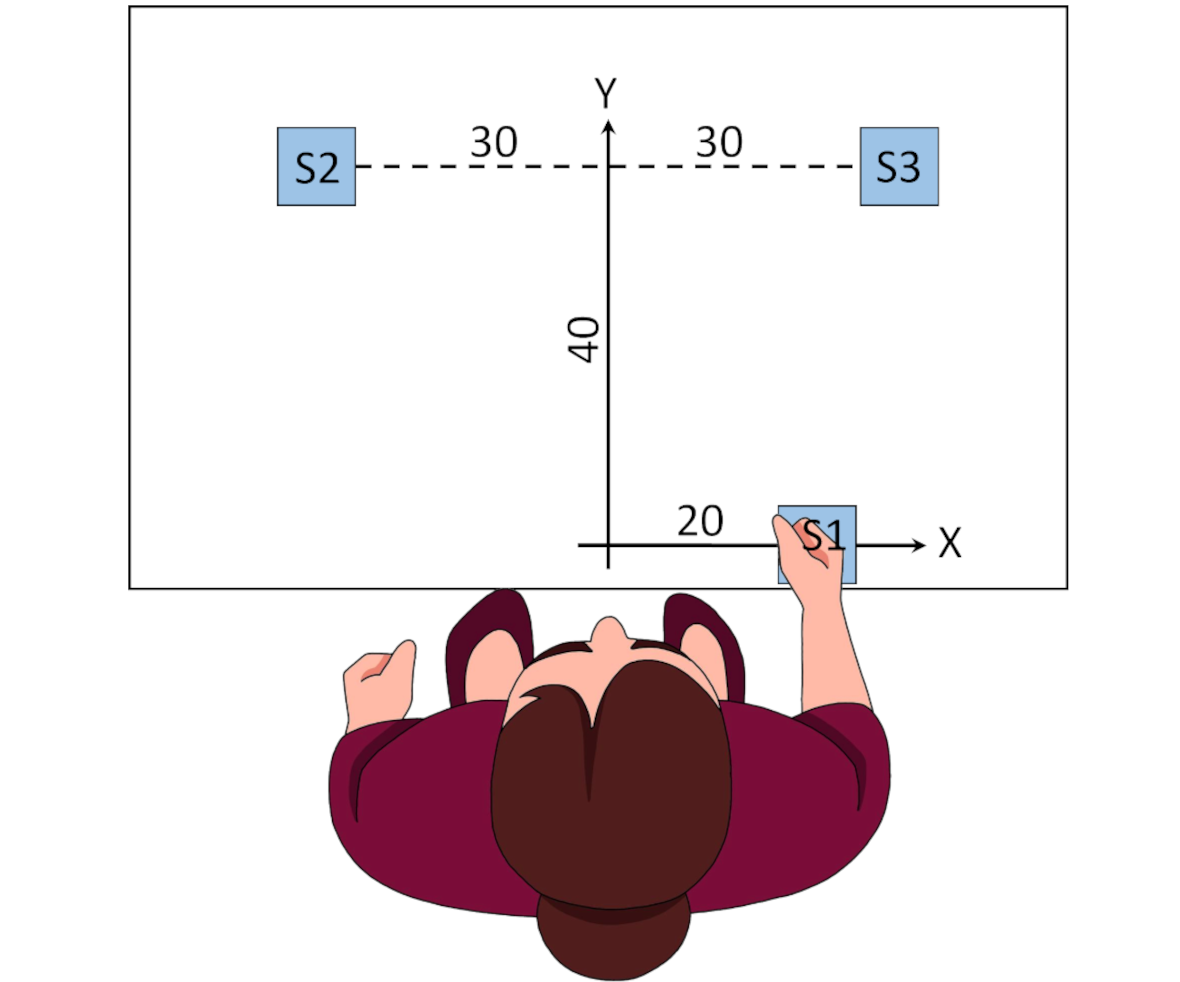}
		\label{fig:table-sketch}
	}
	\subfloat[][]{
	    \includegraphics[width=0.23\linewidth]{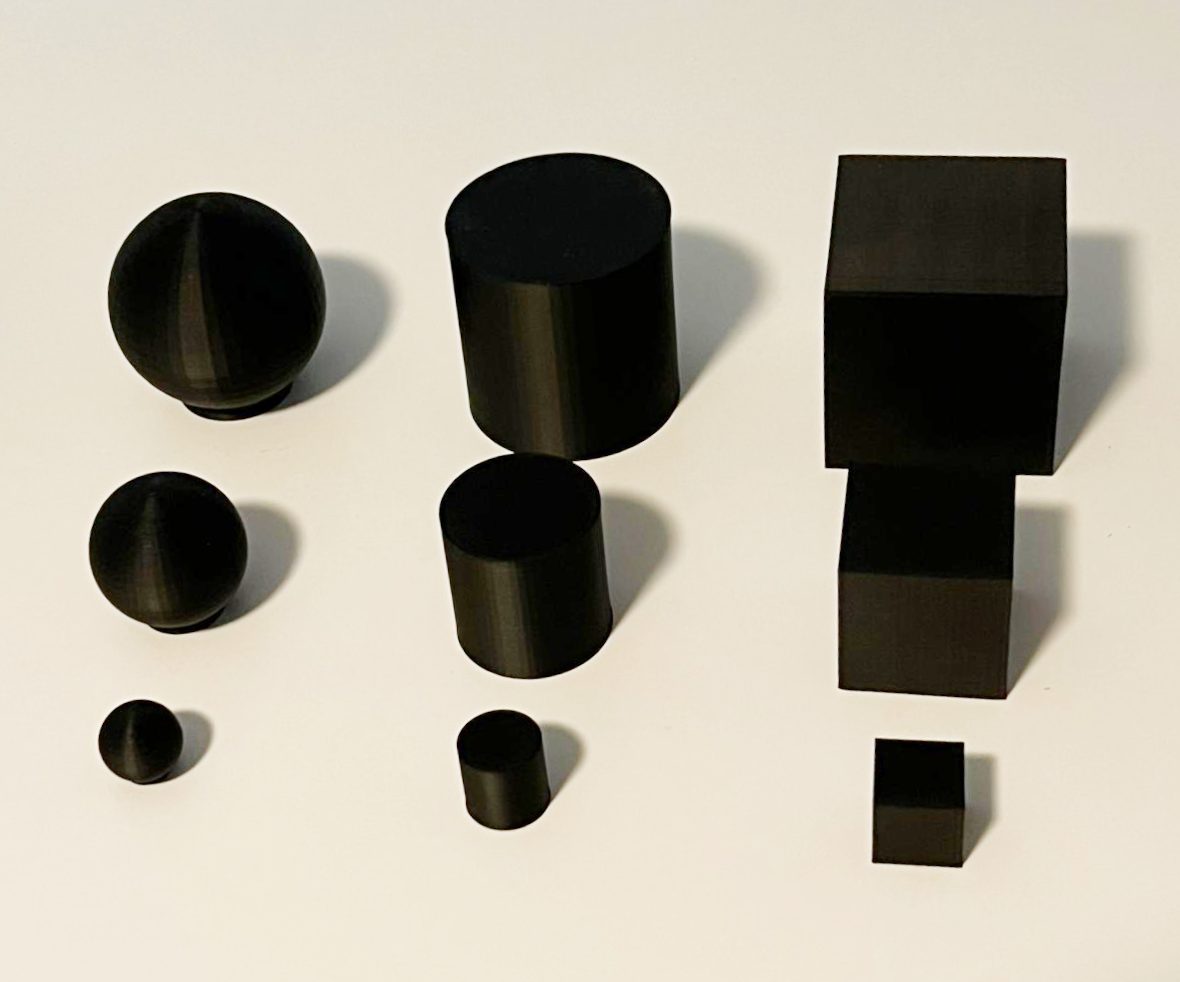}
		\label{fig:so}
	}
	\subfloat[][]{
	    \includegraphics[width=0.23\linewidth]{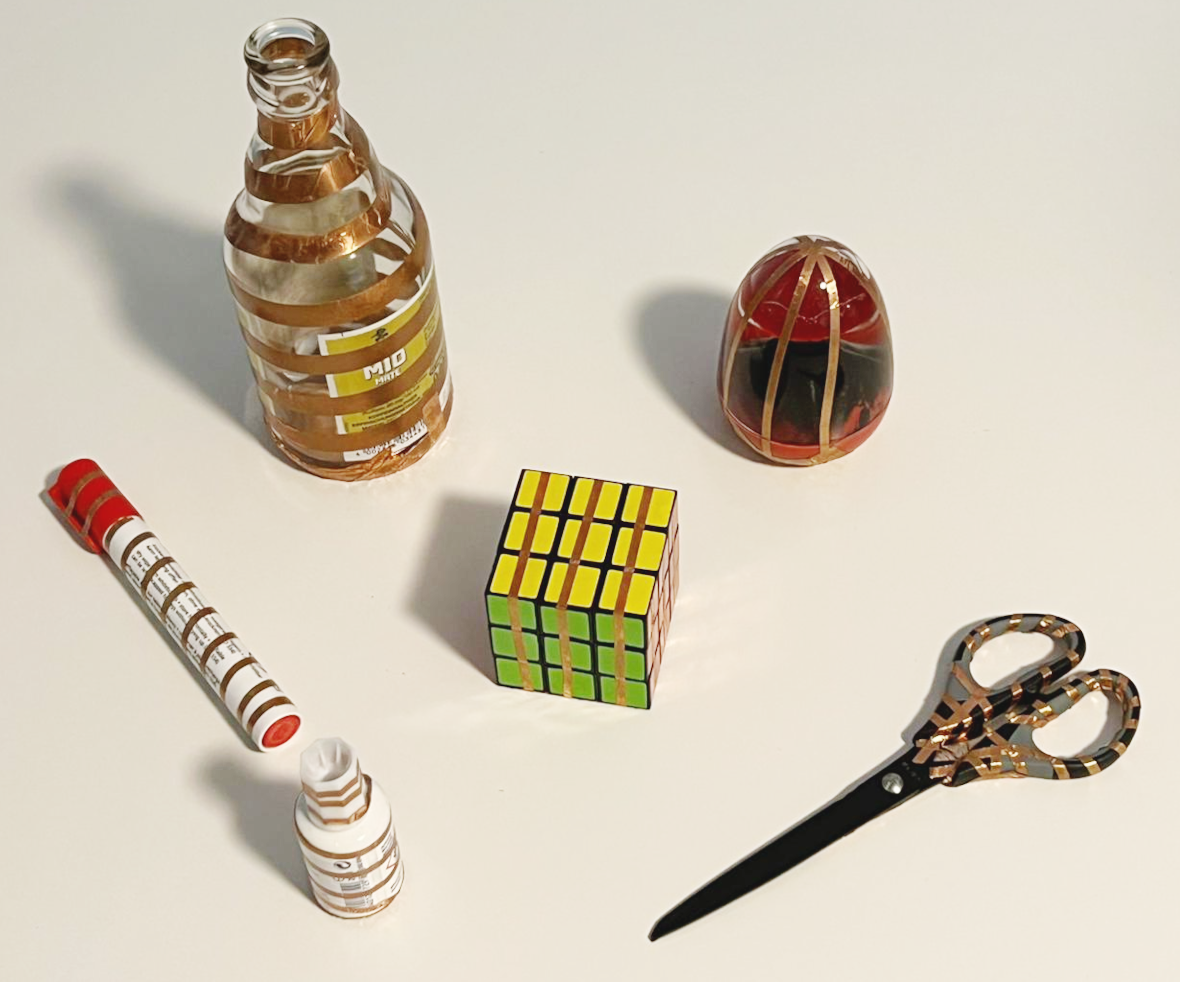}
		\label{fig:ro}
	}
    \caption{Experiment setup: (a) Test subject while participating in the experiment. (b) Experiment setup with the marked hand resting position and touch sensor $S1$, the object position and sensor $S2$, and the target position and sensor $S3$. A photograph of (c) the synthetic and (d) the real objects used in the experiment - \emph{pen, glue, bottle, Rubik's cube, egg-vulcano, toy,} and \emph{scissor}.}
    \Description{The figure shows the experiment setup in four subfigures: (a) Photo of a test subject while participating in the experiment. (b) Sketch of the experiment setup with a marked hand resting position and touch sensor $S1$, the object position and sensor $S2$, and the target position and sensor $S3$. (c) A photograph of the synthetic objects. (d) A photograph of the real objects used in the experiment - \emph{pen, glue, bottle, Rubik's cube, egg-vulcano, toy,} and \emph{scissors}.}
    \label{fig:setup}
\end{figure*}
In this work, we have selected to use LSTM-NN since they have already proven their utility for handling multivariate time series in numerous application fields~\cite{DBLP:journals/corr/SakSB14, DBLP:journals/corr/abs-1801-07962, PHAM2017218}.
The networks' architecture is depicted in Figure~\ref{fig:nn}.
The LSTM layer used a \emph{tanh} state and \emph{sigmoid} gate activation functions and was linearly connected to the subsequent fully connected layer FC.
To avoid overfitting, a dropout layer with a rate of $0.2$ was connected between the LSTM and FC layers.
The output layer consisted of a single fully connected neuron with linear activation for the regression networks or of a fully connected layer with \emph{softmax} activation for the classification networks.
To select the sizes of the LSTM and FC layers, we started each evaluation with a small network and progressively increased the number of (hidden) neurons to find the best performance while still avoiding overfitting.
The concrete values are presented in the respective sections.

In all cases, the training sets were shuffled once at the beginning, and the networks were trained with the Adam optimizer using L2 norm gradient threshold, L2 regularization with $\alpha=10^{-4}$, and mini-batch size of $32$. 
We used the root mean square error as a loss function in the regression and the cross-entropy loss in the classification tasks.

%%----------------------------------------------------------------------------------------------
\section{Data Collection}\label{sec:data-collection}
We collected high-precision hand and finger motion tracking data for R2G and object transport (OT) actions from $16$ adults ($13$ self-identified as male, $3$ as female, mean age $\mu=26.13$, $\sigma=1.78$).
In this data collection study, the participants had to reach out with the right hand and grasp an object placed on a predefined \emph{object position} (R2G task), and then lift the object and move it to a predefined \emph{target position} (OT task), as illustrated in Figure~\ref{fig:setup}.
We used $16$ different objects, and the task was repeated three times for each object, resulting in $48$ data sequences per participant.
All participants were briefed about the purpose of this research, the data privacy policy, and the used materials and methods.  
Each participant was allowed to practice the experiment task until it felt convenient and could take a break or abandon the experiment at any time. 

We used a word-space coordinate system with the $xy$-plane parallel to the participant's transverse plane and the $yz$-plane parallel to the sagittal plane~(see Figure~\ref{fig:table-sketch}).
The \emph{hand resting position} was chosen slightly to the right for convenience, and the \emph{object} and \emph{target} positions were within the convenience range for all participants (Figure~\ref{fig:table-sketch}).
All three positions were touch sensors.

The $16$ study objects were split into two distinct sets - synthetic and ``real'' objects.
The set of synthetic objects is shown in Figure~\ref{fig:so} and consists of three regular geometric solids - sphere, box, and cylinder, each in three different sizes - small ($2$cm), medium ($4$cm) and large ($6$cm). 
The rationale behind the selected shapes is that we wanted to provoke the participants to use similar grasps for each object.
For instance, a higher cylinder would have been easy to distinguish from a sphere since most participants would use a ``cylindrical grasp'' for the first and a ``spherical grasp'' for the second.
With the selected objects, we expected the participants to use a precision grasp in all cases with a different number of fingers, depending on the object's size.
In this case, the discrimination algorithm will need to learn the very subtle kinematic differences of the grips in order to distinguish between objects of the same size but with different shapes. 

The set of real objects, illustrated in Figure~\ref{fig:ro}, consisted of seven objects one would commonly find in an office environment. 
We selected objects with appropriate size, that have well-understandable grasping and usage affordances, and are handheld and commonly moved (e.g., not a puncher or a flowerpot).
The resulting set (cf. Figure~\ref{fig:ro}) is neither complete nor universally generalizable but sufficient to gain some initial insights into the domain.

The synthetic objects were 3D printed with conductive filament, and the real objects were enhanced with copper tape to enable touch transfer to the surfaces. 
The overall setup, although simple, enables high-fidelity motion capturing where each relevant event (e.\,g., when the participant lifted her hand from the start position or touched the object) can be reliably detected. 

%------------------------------------------------------------------------------------------------
\section{Model Training and Evaluation}\label{sec:eval}
From the $16$ participants that took part in the experiment, $768$ tagged motion sequences were collected.
From these, we excluded $5$ paths from the evaluation because of erroneous (de-)activation of the touch surfaces or erroneous task performance (reaching toward the target instead of the object position in one case, excessive performance time of $5$ seconds in another case) that was not noticed by the experimenter. 
The reach-to-grasp (R2G) phase was then extracted from the remaining $763$ sequences.

\begin{figure}[!b]
	\centering
	\includegraphics[width=0.99\linewidth]{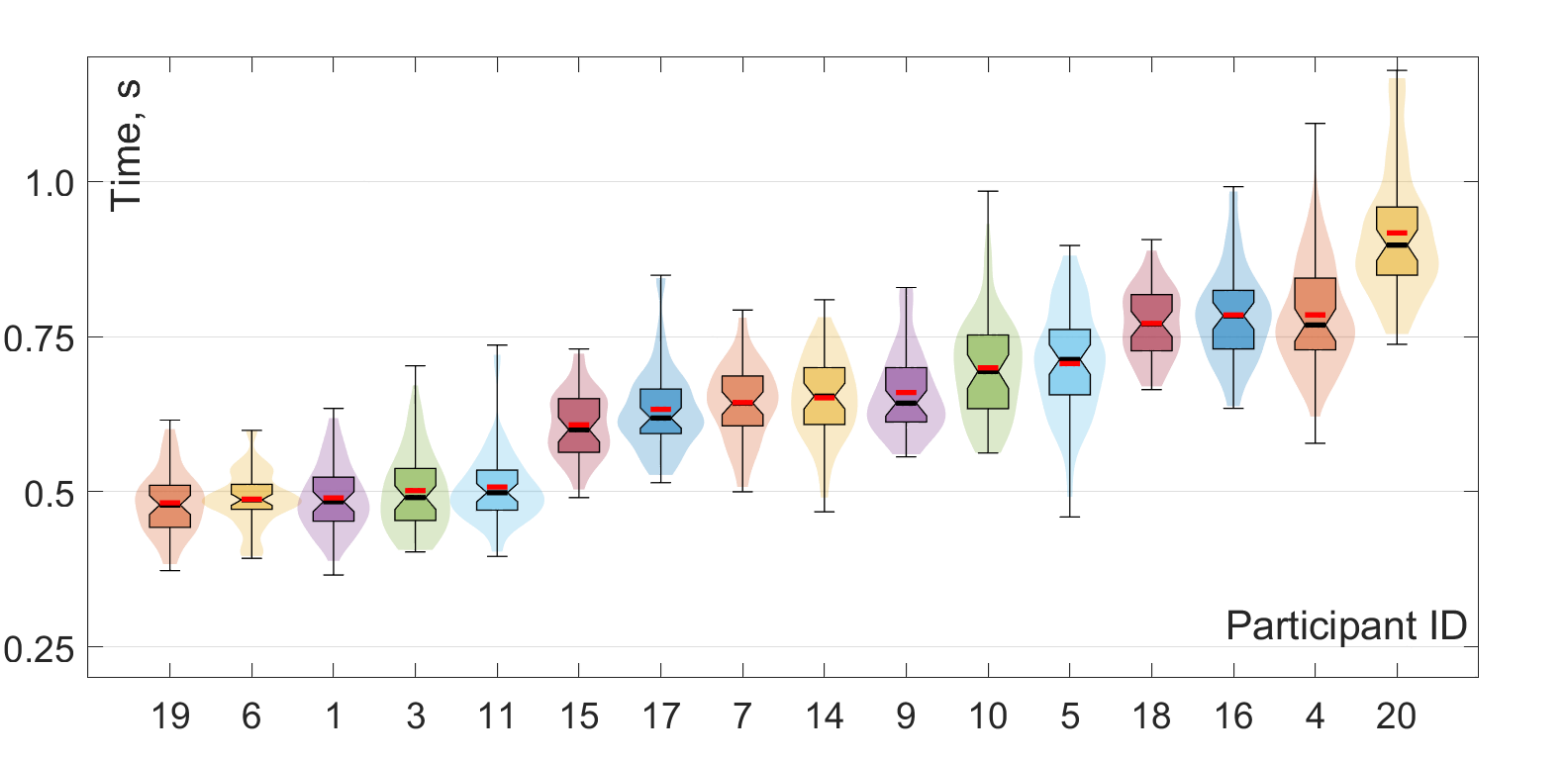}
    \caption{Performance time in seconds of the R2G phase for each participant.}
    \Description{}
    \label{fig:r2g-performance}
\end{figure}

Violin plots of the participants' motion times for the R2G phase are shown in Figure~\ref{fig:r2g-performance}. As one can see in the plots, each participant was relatively consistent in motion velocity and performance time, but there were some significant differences between the participants. Similarly, there was a significant difference in the performance speed for different objects, which corresponds well to Fitts' law.

In the evaluations presented in the following sections, we investigated the LSTM-NN's ability to predict the \emph{time} until the user grasps the intended object, the \emph{distance} to the object, and the target \emph{object} itself. 
Therefore, we constructed three different feature sets from the extracted R2G sequences. The VH set contains only the hand movement velocity. 
The VH+FP feature set extends the VH by adding the 15-dimensional finger polygon features. 
Finally, the VH+FP+PP set contains all $31$ features extracted from the sensors (see Section \ref{sec:features}). 
The R2G feature sequences were then split into small subsequences of a fixed length of either $25$, $50$, or $75$ samples. 
To balance the number of data points in the different sets, we added some overlapping subsequences, which resulted in approximately $35.000$ data points for each subsequence length. 
For the 4-fold cross-validation, the respective dataset was split into four equal parts. In four consecutive training/evaluation sessions, one part was used for validation, and the remaining $75$\% was used for training.  

\subsection{Time until Grasp and Distance to the Target}
\begin{figure}[!t]
\centering
\begin{minipage}[t]{.45\textwidth}
	\centering
	\includegraphics[width=1.0\linewidth]{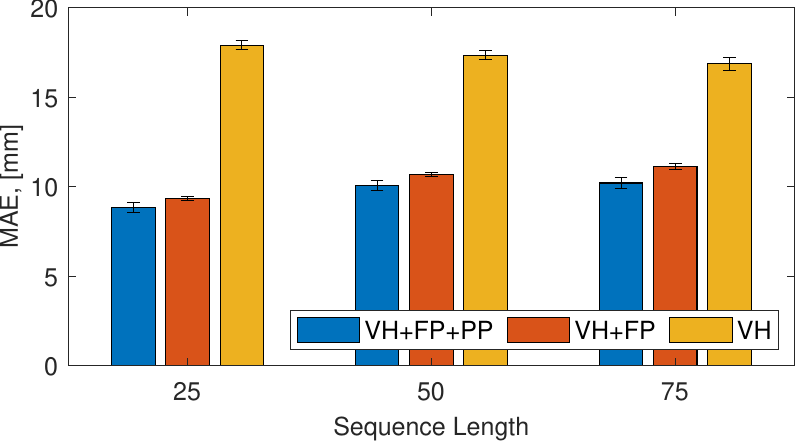}
    \captionof{figure}{Mean Average Error (MAE) in millimeters of the LSTM-NN when trained to predict the distance between the current hand position and the intended target object for different sequence lengths and feature sets. The whiskers show the $\pm$ standard deviation.}%
    \label{fig:mae-distance}
\end{minipage}%
\hspace{0.1\textwidth}%
\begin{minipage}[t]{.45\textwidth}
	\centering
    \includegraphics[width=1.0\linewidth]{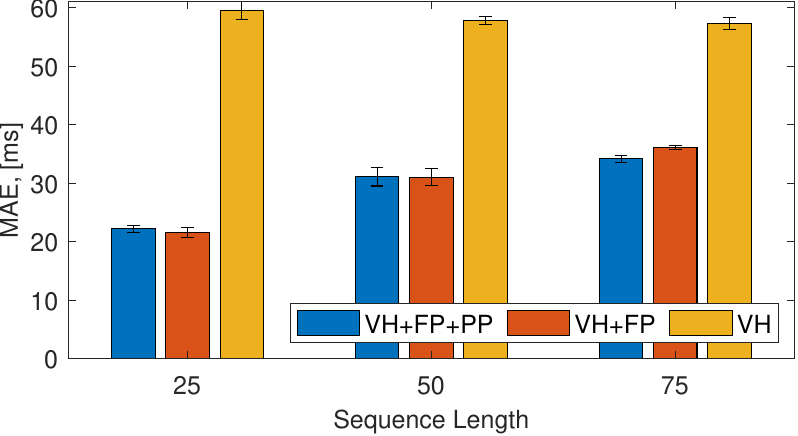}
    \captionof{figure}{Mean Average Error (MAE) in milliseconds of the LSTM-NN when trained to predict the time until the intended object is grasped. The whiskers show the $\pm$ standard deviation.}
    \Description{}
    \label{fig:mae-time}
\end{minipage}%
\end{figure}
In this evaluation, we wanted to test which constellation of local hand features and sequence lengths will allow us to extract precise information about (a) the moment the user grasps the object and (b) the current distance between the target object and the user's hand. 
The network was configured with $64$ hidden neurons in the LSTM layer and $16$ neurons in the FC layer (24k-37k trainable parameters) and was trained for $60$ epochs with a learning rate $\eta=1\cdot10^{-3}$ and minibatch size of $32$. The tests were conducted with a 4-fold cross-validation.

Figure~\ref{fig:mae-distance} shows the mean average error (MAE) for the networks trained to estimate the distance to the target with the different combinations of sequence lengths and features. 
The precision of the networks predicting the time until grasp is shown in Figure~\ref{fig:mae-time}.

Overall, the networks achieved remarkable accuracy in all cases. 
The distance estimations with even the simplest, one-dimensional VH feature (cf. Figure~\ref{fig:mae-distance}) had MAE of less than $18$ mm, comparable with those reported in the related work~\cite{unscripted, Xia:2014, unscr35}, despite using a considerably less sophisticated approach. 
Similarly, the time to grasp estimations with the same simple feature set had MAE better than $60$ ms. 
When the FP is added to the mix, the accuracy increases significantly\footnote{Strictly, training and testing NNs is a stochastic process, and the differences between performance metrics must be evaluated with inferential statistics. Nevertheless, this is highly uncommon in the literature and--given the well-separated, normal distributions--usually unneeded. Thus, we omit reporting these here for clarity but have verified all ``significant'' differences.}. 
In contrast, the PP feature set does not seem to lead to any noteworthy further improvement.
The best performance in the distance estimation was achieved with the 25-sample sequence and the VH+FP+PP feature set (mean MAE $\mu = 8.84$ mm, standard deviation $\sigma = 0.26$ mm), followed by VH+FP for the same sequence length ($\mu=9.35$ mm, $\sigma=0.13$ mm). 
For the time to grasp estimation, the 25-sample sequence and the VH+FP feature set performed best ($\mu=21.53$ ms, $\sigma=0.85$ ms), followed by the VH+FP+PP features ($\mu=22.53$ ms, $\sigma=0.55$ ms).
\begin{figure}[!t]
	\centering
 	\subfloat{
    	\includegraphics[width=0.97\linewidth]{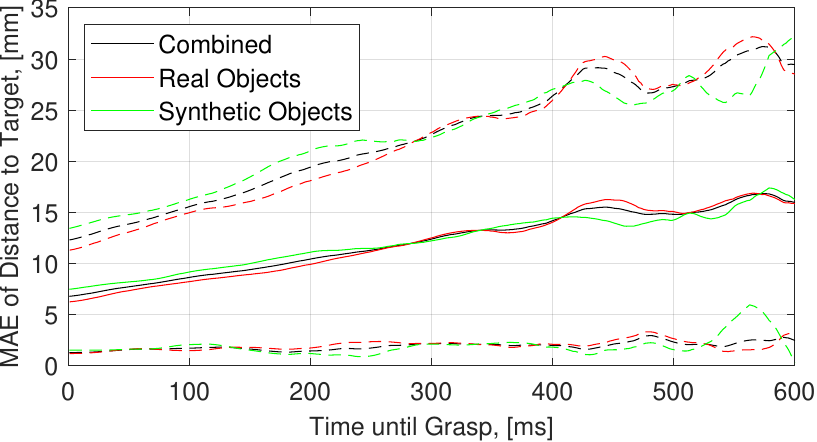}
        \label{fig:distance-vs-time}
    }\\
    \subfloat{
        \includegraphics[width=0.97\linewidth]{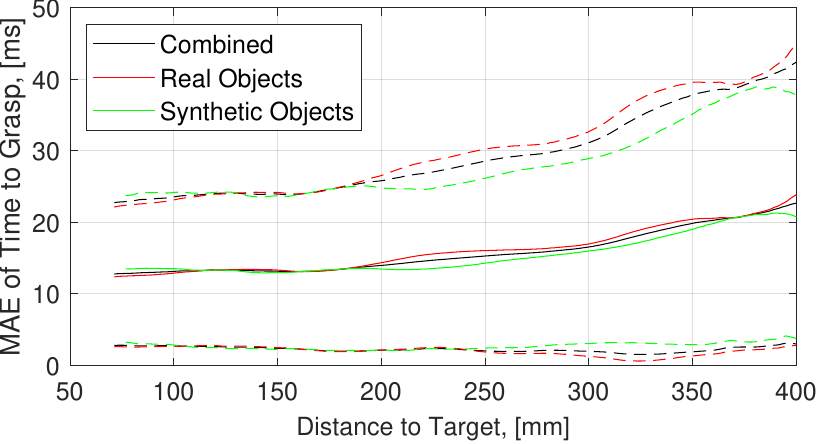}
        \label{fig:time-vs-distance}
    }
    \caption{Mean Average Error in \emph{(top)} the distance to target estimations w.r.t. the (measured) time until the user grasps the target and \emph{(bottom)} the time to grasp estimations w.r.t. the distance between the object and the hand}.
    \Description{}
    \label{fig:simulation-distancetime}
\end{figure}
As the hand approaches the intended object, one would expect that the uncertainty in the prediction decreases. 
In some cases, this can be enforced by training the model using, e.\,g., the Mean Absolute Percentage Error loss function, but this comes at the cost of decreased overall performance, especially when the hand is further away from the object. 
Instead, we decided to use RMSE as a loss function to optimize the overall performance with the 25-sample sequence and VH+FP feature set. 
We then used simulations to gain insights into the model's behavior with regard to time and distance to the target. 
The results of these simulations are shown in Figure~\ref{fig:simulation-distancetime}. 
Figure~\ref{fig:distance-vs-time} depicts how the imprecision in the distance estimations decreases with the time remaining until the user grasps the object. 
In particular, the mean error rate falls under $15$ mm already $400$ ms before the grasp. 
In addition, there is no observable difference between real and synthetic objects. 
The error rates of the time estimations show a similar pattern, with the mean error under $20$ ms as the hand is closer than $35$ cm from the object and under $15$ ms as it comes closer than $20$ cm. 

Overall, given the considerably simpler tracking setup of the VH+FP features (one only needs to track the user's fingertips) and the negligible performance decrease of less than one millimeter, we consider this feature set in combination with the 25-sample sequence to be the best choice for the task. 
We merged the two regression networks into one to further decrease the algorithm's complexity. 
Therefore, we only added one additional neuron to the output layer while keeping all other layers unchanged. 
This network was then trained to simultaneously predict the time until grasp and the distance to the target, and achieved results quantitatively and qualitatively indistinguishable from the two independent networks. 

%------------------------------------------------------------------------------------------------
\subsection{Prediction of Target's Size and Shape}
\begin{figure}[!t]
	\centering
 	\subfloat{
    	\includegraphics[width=0.98\linewidth]{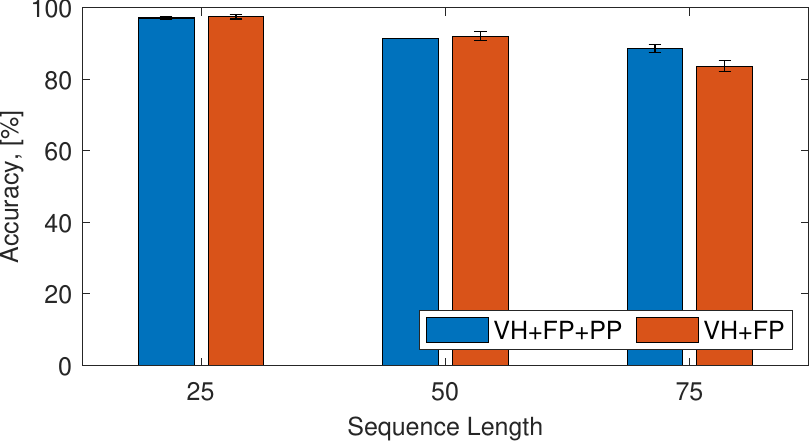}
    }\\
    \subfloat{
        \includegraphics[width=0.98\linewidth]{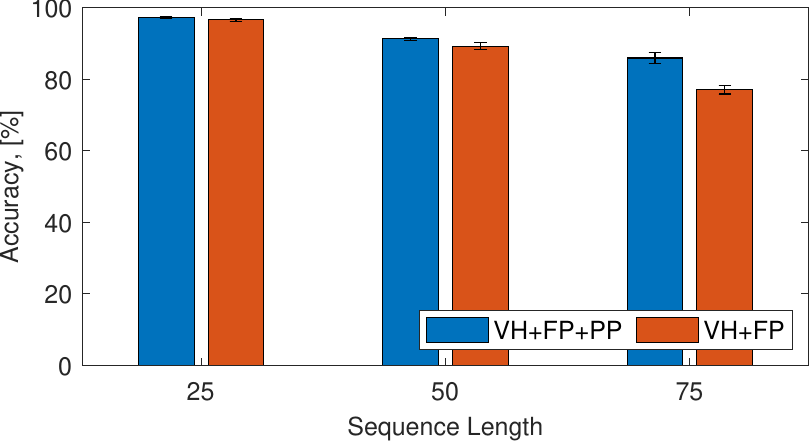}
    }
    \caption{Accuracy in percentages of the discrimination of the \emph{(top)} real and \emph{(bottom)} synthetic objects for different combinations of features and sequence lengths.}\Description{}
    \label{fig:objects}
\end{figure}
\begin{figure}[!t]
	\centering
 	\subfloat{
    	\includegraphics[width=0.98\linewidth]{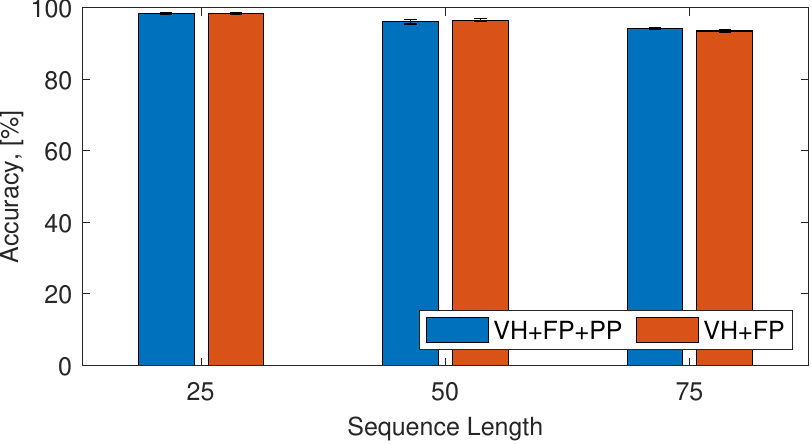}
    }\\
    \subfloat{
        \includegraphics[width=0.98\linewidth]{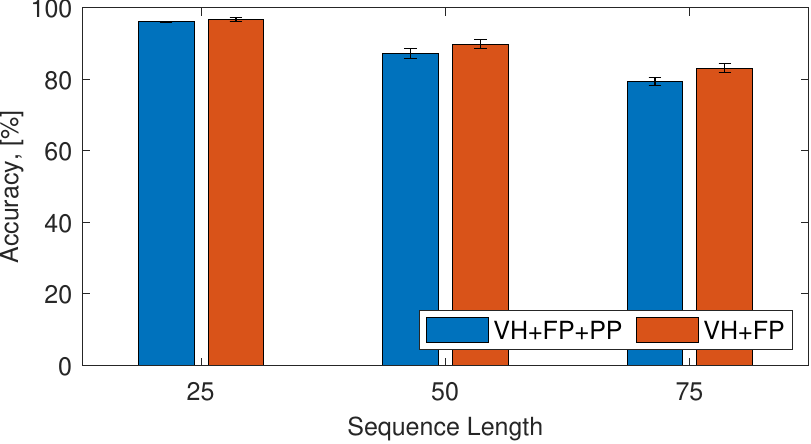}
    }
    \caption{Accuracy in percentages of the discrimination of the synthetic object's \emph{(top)} size and \emph{(bottom)} shape for different combinations of features and sequence lengths.}
    \Description{}
    \label{fig:sizeshape}
\end{figure}
\begin{figure}[!t]
	\centering
    \includegraphics[width=0.98\linewidth]{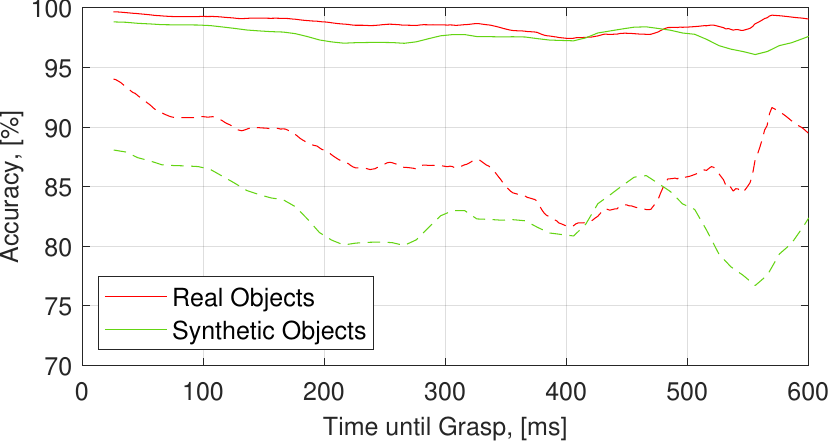}
    \caption{Accuracy of the object prediction w.r.t. the time until the user grasps the target. The accuracy for the real objects is depicted by the red lines, and for the synthetic objects by the green lines. The solid lines represent the mean accuracy, and the dashed lines represent the accuracy of one standard deviation below the mean.}
    \Description{}
    \label{fig:accuracy-vs-time}
\end{figure}
In this evaluation, we focused on assessing the network's capability to recognize the object the user intends to grasp.
We used two sets of objects: \emph{real} and \emph{synthetic}. 
For the synthetic objects, we further divided the evaluation into two distinct aspects: recognizing the object's \emph{size} and \emph{shape}. 
The network was configured with $128$ hidden neurons in the LSTM layer and $16$ neurons in the FC layer.

Figure~\ref{fig:objects} shows the accuracy of the networks trained with different combinations of sequence lengths and features to predict the intended object. 
The accuracy of the networks predicting the \emph{size} and \emph{shape} of the synthetic objects is shown in Figure~\ref{fig:sizeshape}. 
Here we only tested the feature combinations VH+FP and VH+FP+PP since the hand motion velocity alone does not provide sufficient information about the to-be-grasped object. In particular, the best accuracy we could achieve with this feature set was $\approx 30\%$, for all tested hyperparameters. 

As one can see in the figures, the 25-sample sequence significantly outperformed the other configurations. 
In contrast, the PP features did not add any substantial improvement. 
Interestingly, the inclusion of the PP features in the shape discrimination task decreased the overall accuracy instead of improving it. 
One reason for this might be that we did not compensate for the users' different finger or hand sizes. 
Indeed, while the fingertips' motions are only correlated to the object's size and shape, the movements of the proximal phalanges also depend on the hand size. 
In contrast, the network seems able to abstract from this information in the (simpler) size discrimination task.

To explore the network's behavior as the hand approaches the intended object, we again used the VH+FP network trained with $75\%$ of the 25-sample sequences and simulated the run-time input with the remaining data. 
The results are shown in Figure~\ref{fig:accuracy-vs-time}.  
The results reveal the same decreasing trend, although not as pronounced as with the regression networks. 
The mean accuracy for both real and synthetic objects remained above $95\%$ in all cases. 
It gradually increased in mean and spread in the last $400$ ms before the grasp.  
Considering the conservative boundary of one standard deviation below the mean (dashed lines in Figure~\ref{fig:accuracy-vs-time}), one can see that the synthetic objects were more challenging for our algorithm. 
Until the last $200$ ms before the grasp, this worst-case accuracy for discriminating synthetic objects was approximately $80\%$ (98\% in the average case). 
In contrast, the real objects showed a higher discriminability, with worst-case accuracy reaching at least $85\%$ as early as $350$ ms before the actual grasp. The average case accuracy at that point was above 97\%. 

For the discrimination task, the 25-sample sequence combined with the VH+FP feature set is again the most effective choice. The VH features alone were insufficient for discrimination of the future object, even after extensive hyperparameter tweaking, and including the PP features did not aid the discrimination accuracy.

%%------------------------------------------------------------------------------------------------
\subsection{Generalization for Unknown Users}
The findings presented in the previous two sections give an overview of how the network would perform if the user is known or when the training set contains sufficient data to capture the entire between-user variability. Unfortunately, this is not feasible in most cases. Therefore, we conducted leave-one-user-out (L1UO) tests to assess the networks' ability to generalize for unknown users and transfer-learning tests to evaluate their ability to adapt to new users quickly. 
In the L1UO tests, the LSTM was trained with the 25-sample VH+FP sequences of all but one user and tested with the data of the left-out user.
We also conducted leave-one-session-out and leave-one-object-out (for the synthetic objects) tests, but found only a marginal drop in the overall performance, with no apparent influence of the left-out object or session. Thus, we omit these here for brevity. 

In the transfer-learning tests, the (already trained) LSTM was trained further for 50 epochs with a few sequences from the left-out user. This technique aimed to simulate a real-world scenario where tracking data is collected during system usage and utilized for user-specific adaptation. On the other hand, it can also be used to assess the network's performance as new data is added in the future. We tested the effect of adding 50, 150, and 250 additional samples. Please note that 250 new sequences could be acquired from only one grasp gesture. Nevertheless, more trials would better capture the user's specifics.
\begin{figure}
\centering
\begin{minipage}[t]{.45\textwidth}
	\centering
	\subfloat{
	    \includegraphics[width=0.29\linewidth]{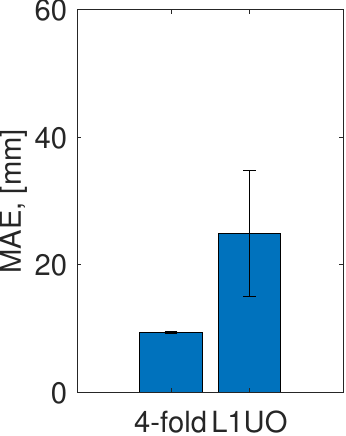}
	}
	\subfloat{
	    \includegraphics[width=0.65\linewidth]{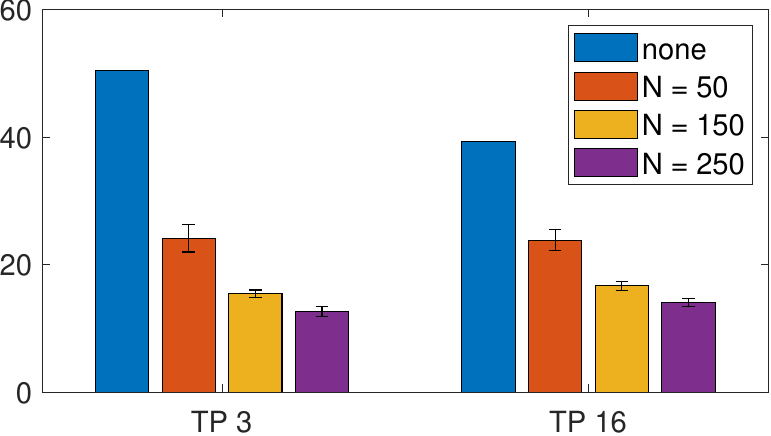}
	}\\
 	\subfloat{
	    \includegraphics[width=0.29\linewidth]{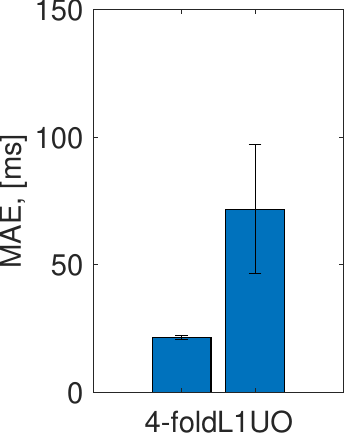}
	}
	\subfloat{
	    \includegraphics[width=0.65\linewidth]{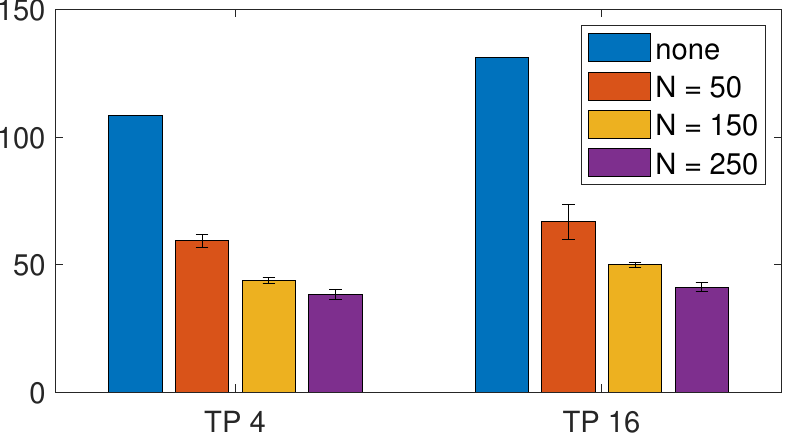}
	}
    \captionof{figure}{Mean Average Error (MAE) for the \emph{(top)}~distance to target and \emph{(bottom)}~time until grasp estimations with the 25-sample VH+FP network. \emph{(left)}~Comparision of the results from the 4-fold cross-validation and the L1UO test. \emph{(right)}~Results of the transfer learning for the two participants for whom the networks performed the worst.}
    \Description{}
    \label{fig:l1uo-regression}
\end{minipage}%
\hspace{0.05\textwidth}%
\begin{minipage}[t]{.45\textwidth}
	\centering
	\subfloat{
	    \includegraphics[width=0.29\linewidth]{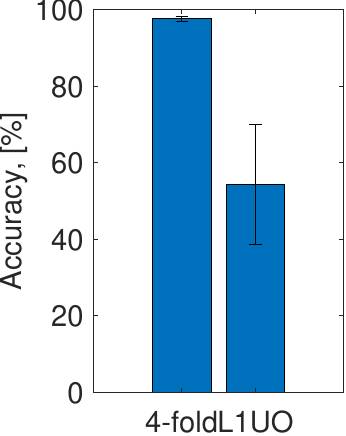}
	}
	\subfloat{
	    \includegraphics[width=0.65\linewidth]{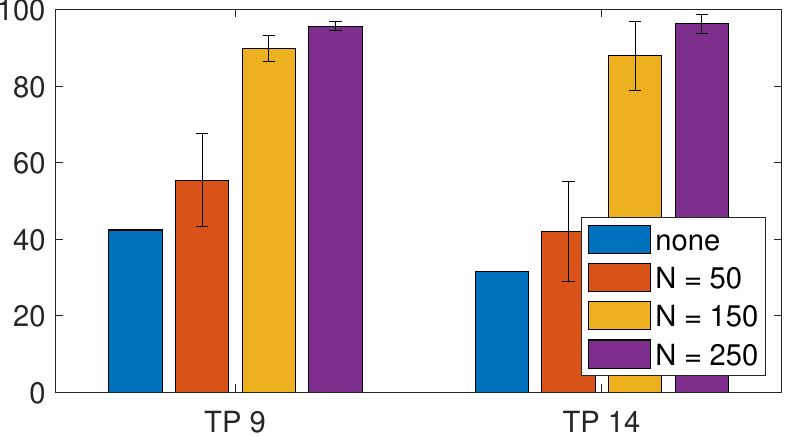}
	}\\
 	\subfloat{
	    \includegraphics[width=0.29\linewidth]{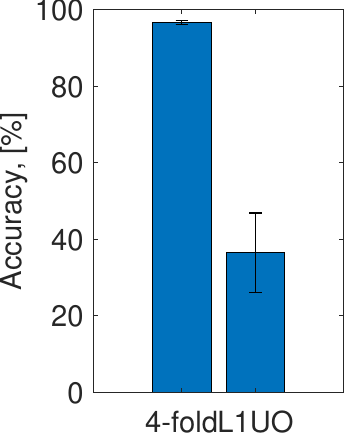}
	}
	\subfloat{
	    \includegraphics[width=0.65\linewidth]{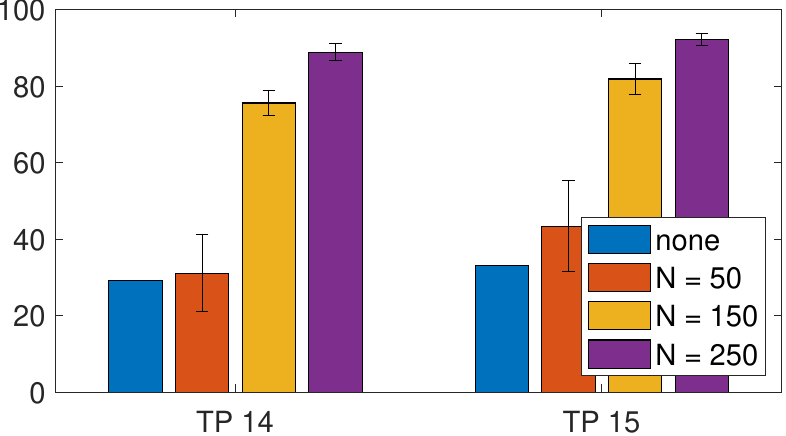}
	}\\
 	\subfloat{
	    \includegraphics[width=0.29\linewidth]{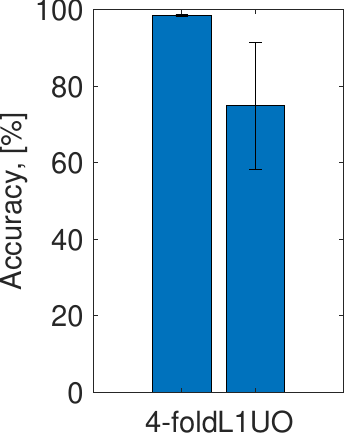}
	}
	\subfloat{
	    \includegraphics[width=0.65\linewidth]{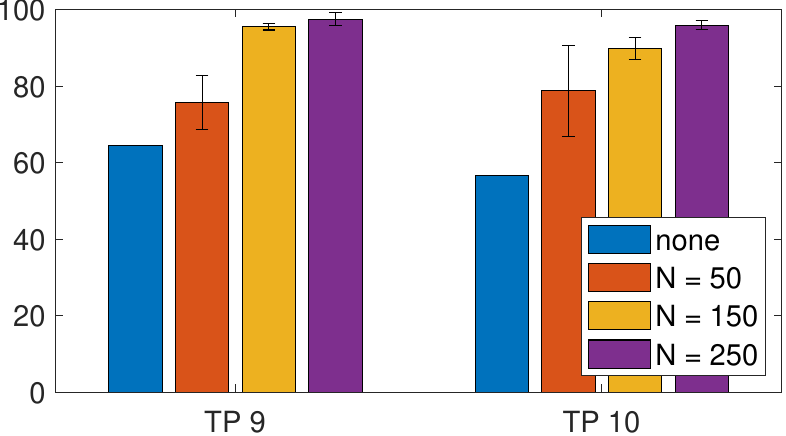}
	}
    \captionof{figure}{Accuracy for the discrimination of the \emph{(top)}~real objects, \emph{(middle)}~ synthetic objects, and \emph{(bottom)}~the object's size with the 25-sample VH+FP network. \emph{(left)}~Comparision of the results from the 4-fold cross-validation and the L1UO test. \emph{(right)}~Results of the transfer learning test performed for the two participants for whom the networks performed the worst.}
    \Description{}
    \label{fig:l1uo-objects}
\end{minipage}%
\end{figure}

Figure~\ref{fig:l1uo-regression} shows the results of the L1UO and the transfer-learning test for the regression network, trained with the 25-sample sequence and VH+FP feature set. In the distance estimation task, the mean error rate increased significantly to $24.89$ ($\sigma=9.91$) millimeters and up to $51.87$ mm in the worst case for test participant TP 3. While this is a considerable drop in performance, it may still be within the usable range for some applications. Furthermore, the network can quickly adapt to unknown users (see Figure~\ref{fig:l1uo-regression} left). Training it further with only $50$ new data points (it takes less than a second on a Core i7 CPU) decreases the mean error rate to under $15.5$ mm, even for the participants for whom it performed worse (TP 3 and TP 16). The time-until-grasp estimations show the same qualitative behavior. 
The mean L1UO error rate of $71.78$ ($\sigma=25.30$) milliseconds is considerable, but transfer learning helps again, with less than $150$ new data points needed to achieve error rates under $50$ ms in the mean.

For the discrimination of the to-be-grasped object, the networks reveal similar behavior~(see Figure~\ref{fig:l1uo-objects}), with one substantial difference -- for most users, the initially pre-trained network would be barely usable without adaptation. 
This is not surprising, considering that the network needed to discriminate both the object and the way the user grasps it. In particular, in agreement with Feix et al.~\cite{Feix2014graspobject}, different users grabbed the same object with different grips and in different positions. Given the small number of test users and the variety of grips and object shapes, it was difficult for the network to generalize to data never seen before. 
Notably, specialization of the network with $150$ user-dependent data points increases the accuracy above $87$ \% for the real and $75$ \% for the synthetic objects, as illustrated in Figure~\ref{fig:l1uo-objects} \emph{top} and \emph{bottom}. Thus, we can reasonably expect that the generalization for users never seen before will significantly improve with a larger dataset. 

In contrast, the size discrimination network can be used directly with a minor accuracy penalty for unknown users.
This performance drop is quickly alleviated by the user-specific adaptation (s. Figure~\ref{fig:l1uo-objects} \emph{bottom}).

Overall, the results of the L1UO tests indicated that the distance, time, and size estimations generalize well for unknown users and can be easily improved by specializing the network for the particular user, which can be performed in the background with data collected on the fly. 
Precise object discrimination is, in contrast, only available after user calibration. Nevertheless, there is a strong indication that these will improve with a larger dataset that captures the variety of grasps used for the same object.  

%%------------------------------------------------------------------------------------------------
\section{Discussion and Future Work}
The results presented in this work demonstrate how one can determine the distance to the to-be-grasped object with high precision, the approximate time until it is grasped, and at least the object size, using only local hand dynamics. 
The presented algorithms also showed promising results for discriminating objects with sufficiently different grasping affordances. 
This makes a tremendous amount of additional information available to the user interface long before the user ever reaches an object. 
Indeed, even the fastest participant needed approximately $300$ ms between the PMV and the first touch of the object. 
The two regression networks were able to deliver high-precision predictions from the very first feature sequence. 
Thus, the interface will still have more than $280$ ms to use the information, even in this exceptional case. 
In the more general case, this time buffer increases to more than $400$ ms. 

The time until grasp could be reliably estimated with an accuracy within the synchronization precision of the dataset (approx. $25$ ms), and the accuracy of the distance estimations reached MAE of under a centimeter for known and just under three centimeters for unknown users. 
Moreover, these accuracy levels were achieved despite the significant difference in users' performance and grasping behavior. 
It is worth also repeating that the features were extracted directly from the marker data, without any user-specific adjustments, e.\,g., for different finger sizes. 

The direct application of this would be to incorporate it in the various psychological models of hand and grasp motion. In these, the acquired user data is usually scaled in the unit time interval $\left[0, 1\right]$, and the modeled trajectories are a function of this relative motion time, which renders them unusable for real-time applications. 
However, with the accurate time-to-grasp estimation provided by our algorithm, the total time of the R2G motion can be easily calculated. This enables the calculation of appropriate time scaling factors, making such psychological models usable in real-time.

Another application of the time-to-grasp prediction would be to reduce the point-to-point latency of the interactive system by providing it with advanced information about the intended object. With this information, the system can decide to, e.g., evoke an action early on, such that the system's reaction is aligned with the moment the user grasps the object.

Although achieving recognition rates comparable to or even better than those reported in the related work~\cite{Xia:2014,Heumer:2008,Vatavu2013a}, our algorithms could not reliably discriminate the object's shape for the synthetic objects. The accuracy increases considerably for the real objects, which have clearly distinguishable grasp affordances. However, without user calibration, it is presumably still insufficient for many applications.
The grasp type seems to be the most decisive factor in this regard, as evident from the shape recognition of the synthetic objects. 
In this case, all objects were deliberately selected with similar grasping affordances, which made them indistinguishable for the algorithm. 
The common confusion of similarly shaped real objects, e.g., glue and the toy, is another indication of this.
Indeed, with our approach, the network needed to recognize not only the target object itself but also \emph{where} and \emph{how} the user will grasp it while abstracting from human factors at the same time.
In contrast, the size discrimination networks easily achieved high recognition rates for both known and unknown users.

One possible solution would be to split the object recognition task into two phases: (a) prediction of the grasp shape and size, and (b) mapping the grip to a target object. 
As already demonstrated, the size can be reliably predicted.
Based on the results for the real objects, early prediction of the class of the grasp also seems plausible. 
Nevertheless, we need a new dataset in which the objects are selected w.r.t all common grasp types, which will be the subject of future experiments.
Another promising direction for future work is the definition of feature sets that are less dependent on hand or finger sizes. 
Trivial candidates are the finger joint angles or the distances between the fingertips. 

Overall, the results already provide a solid working base for practical applications. 
For instance, the target distance prediction enables a rough estimation of the object's position. 
While we did not use the global hand position, it will undoubtedly be known to the interaction system. 
Thus, using the moving hand position as a center and the estimated (undirected) distance as a radius will allow us to find the approximate object position as an intersection of spheres. 
The additional information about the object's size can further reduce the number of potential target candidates. 
A virtual environment -- fully controlled by the designer -- can be constructed such that this information is already sufficient. 
For instance, one can group items with different sizes and grasp affordances together while keeping similarly sized objects further apart. 
The time estimations, albeit unprecise for some users, can further enable dynamic and adaptive interfaces. 
Nevertheless, the presented results are still preliminary and unveil several challenges that will be addressed in future research.

\begin{acks}
We want to thank Valeria for helping design and shape this paper.
This project was partially funded by the Deutsche Forschungsgemeinschaft (DFG, German Research Foundation) – Project 436291335.
%... and thanks to Doug for all the fish.
\end{acks}

\bibliographystyle{ACM-Reference-Format}
\bibliography{bibliography}


\begin{thebibliography}{50}

%%% the Web address.

\ifx \showCODEN    \undefined \def \showCODEN     #1{\unskip}     \fi
\ifx \showDOI      \undefined \def \showDOI       #1{#1}\fi
\ifx \showISBNx    \undefined \def \showISBNx     #1{\unskip}     \fi
\ifx \showISBNxiii \undefined \def \showISBNxiii  #1{\unskip}     \fi
\ifx \showISSN     \undefined \def \showISSN      #1{\unskip}     \fi
\ifx \showLCCN     \undefined \def \showLCCN      #1{\unskip}     \fi
\ifx \shownote     \undefined \def \shownote      #1{#1}          \fi
\ifx \showarticletitle \undefined \def \showarticletitle #1{#1}   \fi
\ifx \showURL      \undefined \def \showURL       {\relax}        \fi
% invisible to TeX
\providecommand\bibfield[2]{#2}
\providecommand\bibinfo[2]{#2}
\providecommand\natexlab[1]{#1}
\providecommand\showeprint[2][]{arXiv:#2}

\bibitem[A~Jones and J~Lederman(2006)]%
        {Jones:2006}
\bibfield{author}{\bibinfo{person}{L A~Jones} {and} \bibinfo{person}{S J~Lederman}.} \bibinfo{year}{2006}\natexlab{}.
\newblock \bibinfo{booktitle}{\emph{Human Hand Function}}. Vol.~\bibinfo{volume}{32}.
\newblock \bibinfo{publisher}{Oxford University Press}.
\newblock
\showISBNx{0195173155}
\urldef\tempurl%
\url{https://doi.org/10.1093/acprof:oso/9780195173154.001.0001}
\showDOI{\tempurl}


\bibitem[Ahmad et~al\mbox{.}(2018)]%
        {unscr3}
\bibfield{author}{\bibinfo{person}{Bashar~I. Ahmad}, \bibinfo{person}{Chrisminder Hare}, \bibinfo{person}{Harpreet Singh}, \bibinfo{person}{Arber Shabani}, \bibinfo{person}{Briana Lindsay}, \bibinfo{person}{Lee Skrypchuk}, \bibinfo{person}{Patrick Langdon}, {and} \bibinfo{person}{Simon Godsill}.} \bibinfo{year}{2018}\natexlab{}.
\newblock \showarticletitle{Selection Facilitation Schemes for Predictive Touch with Mid-Air Pointing Gestures in Automotive Displays}. In \bibinfo{booktitle}{\emph{Proceedings of the 10th International Conference on Automotive User Interfaces and Interactive Vehicular Applications}} (Toronto, ON, Canada) \emph{(\bibinfo{series}{AutomotiveUI '18})}. \bibinfo{publisher}{Association for Computing Machinery}, \bibinfo{address}{New York, NY, USA}, \bibinfo{pages}{21–32}.
\newblock
\showISBNx{9781450359467}
\urldef\tempurl%
\url{https://doi.org/10.1145/3239060.3239067}
\showDOI{\tempurl}


\bibitem[Altch\'{e} and de~La~Fortelle(2017)]%
        {DBLP:journals/corr/abs-1801-07962}
\bibfield{author}{\bibinfo{person}{Florent Altch\'{e}} {and} \bibinfo{person}{Arnaud de La~Fortelle}.} \bibinfo{year}{2017}\natexlab{}.
\newblock \showarticletitle{An LSTM Network for Highway Trajectory Prediction}. In \bibinfo{booktitle}{\emph{2017 IEEE 20th International Conference on Intelligent Transportation Systems (ITSC)}} (Yokohama, Japan). \bibinfo{publisher}{IEEE Press}, \bibinfo{pages}{353–359}.
\newblock
\urldef\tempurl%
\url{https://doi.org/10.1109/ITSC.2017.8317913}
\showDOI{\tempurl}


\bibitem[Ansuini et~al\mbox{.}(2015)]%
        {Ansuini:2015}
\bibfield{author}{\bibinfo{person}{Caterina Ansuini}, \bibinfo{person}{Andrea Cavallo}, \bibinfo{person}{Atesh Koul}, \bibinfo{person}{Marco Jacono}, \bibinfo{person}{Yuan Yang}, {and} \bibinfo{person}{Cristina Becchio}.} \bibinfo{year}{2015}\natexlab{}.
\newblock \showarticletitle{Predicting Object Size from Hand Kinematics: A Temporal Perspective}.
\newblock \bibinfo{journal}{\emph{PLOS ONE}} \bibinfo{volume}{10}, \bibinfo{number}{3} (\bibinfo{date}{03} \bibinfo{year}{2015}), \bibinfo{pages}{1--13}.
\newblock
\urldef\tempurl%
\url{https://doi.org/10.1371/journal.pone.0120432}
\showDOI{\tempurl}


\bibitem[Ansuini et~al\mbox{.}(2006)]%
        {Ansuini:2006}
\bibfield{author}{\bibinfo{person}{Caterina Ansuini}, \bibinfo{person}{Marco Santello}, \bibinfo{person}{Stefano Massaccesi}, {and} \bibinfo{person}{Umberto Castiello}.} \bibinfo{year}{2006}\natexlab{}.
\newblock \showarticletitle{Effects of end-goal on hand shaping}.
\newblock \bibinfo{journal}{\emph{Journal of neurophysiology}} \bibinfo{volume}{95}, \bibinfo{number}{4} (\bibinfo{year}{2006}), \bibinfo{pages}{2456--2465}.
\newblock
\showISSN{0022-3077}
\urldef\tempurl%
\url{https://doi.org/10.1152/jn.01107.2005}
\showDOI{\tempurl}


\bibitem[Asano et~al\mbox{.}(2005)]%
        {unscr7}
\bibfield{author}{\bibinfo{person}{Takeshi Asano}, \bibinfo{person}{Ehud Sharlin}, \bibinfo{person}{Yoshifumi Kitamura}, \bibinfo{person}{Kazuki Takashima}, {and} \bibinfo{person}{Fumio Kishino}.} \bibinfo{year}{2005}\natexlab{}.
\newblock \showarticletitle{Predictive Interaction Using the Delphian Desktop}. In \bibinfo{booktitle}{\emph{Proceedings of the 18th Annual ACM Symposium on User Interface Software and Technology}} (Seattle, WA, USA) \emph{(\bibinfo{series}{UIST '05})}. \bibinfo{publisher}{Association for Computing Machinery}, \bibinfo{address}{New York, NY, USA}, \bibinfo{pages}{133–141}.
\newblock
\showISBNx{1595932712}
\urldef\tempurl%
\url{https://doi.org/10.1145/1095034.1095058}
\showDOI{\tempurl}


\bibitem[Azmandian et~al\mbox{.}(2016)]%
        {Azmandian:2016:hapticRetargeting}
\bibfield{author}{\bibinfo{person}{Mahdi Azmandian}, \bibinfo{person}{Mark Hancock}, \bibinfo{person}{Hrvoje Benko}, \bibinfo{person}{Eyal Ofek}, {and} \bibinfo{person}{Andrew~D. Wilson}.} \bibinfo{year}{2016}\natexlab{}.
\newblock \showarticletitle{Haptic Retargeting: Dynamic Repurposing of Passive Haptics for Enhanced Virtual Reality Experiences}. In \bibinfo{booktitle}{\emph{Proceedings of the 2016 CHI Conference on Human Factors in Computing Systems}} (San Jose, California, USA) \emph{(\bibinfo{series}{CHI '16})}. \bibinfo{publisher}{Association for Computing Machinery}, \bibinfo{address}{New York, NY, USA}, \bibinfo{pages}{1968–1979}.
\newblock
\showISBNx{9781450333627}
\urldef\tempurl%
\url{https://doi.org/10.1145/2858036.2858226}
\showDOI{\tempurl}


\bibitem[Betti et~al\mbox{.}(2018)]%
        {Betti:2018}
\bibfield{author}{\bibinfo{person}{Sonia Betti}, \bibinfo{person}{Giovanni Zani}, \bibinfo{person}{Silvia Guerra}, \bibinfo{person}{Umberto Castiello}, {and} \bibinfo{person}{Luisa Sartori}.} \bibinfo{year}{2018}\natexlab{}.
\newblock \showarticletitle{Reach-To-Grasp Movements: A Multimodal Techniques Study}.
\newblock \bibinfo{journal}{\emph{Frontiers in Psychology}} \bibinfo{volume}{9}, \bibinfo{number}{2} (\bibinfo{year}{2018}), \bibinfo{pages}{990}.
\newblock


\bibitem[C et~al\mbox{.}(2017)]%
        {Santina:2017}
\bibfield{author}{\bibinfo{person}{Della~Santina C}, \bibinfo{person}{Bianchi M}, \bibinfo{person}{Averta G}, \bibinfo{person}{Ciotti S}, \bibinfo{person}{Arapi V}, \bibinfo{person}{Fani S}, \bibinfo{person}{Battaglia E}, \bibinfo{person}{Catalano MG}, \bibinfo{person}{Santello M}, {and} \bibinfo{person}{Bicchi A.}} \bibinfo{year}{2017}\natexlab{}.
\newblock \showarticletitle{Postural Hand Synergies during Environmental Constraint Exploitation}.
\newblock \bibinfo{journal}{\emph{Frontiers in Neurorobotics}} \bibinfo{volume}{11}, \bibinfo{number}{41} (\bibinfo{year}{2017}).
\newblock
\urldef\tempurl%
\url{https://doi.org/doi: 10.3389/fnbot.2017.00041}
\showDOI{\tempurl}


\bibitem[Castiello(2005)]%
        {Castiello:2005:review}
\bibfield{author}{\bibinfo{person}{Umberto Castiello}.} \bibinfo{year}{2005}\natexlab{}.
\newblock \showarticletitle{The neuroscience of grasping}.
\newblock \bibinfo{journal}{\emph{Nature reviews. Neuroscience}}  \bibinfo{volume}{6} (\bibinfo{date}{10} \bibinfo{year}{2005}), \bibinfo{pages}{726--36}.
\newblock
\urldef\tempurl%
\url{https://doi.org/10.1038/nrn1744}
\showDOI{\tempurl}


\bibitem[Cavallo et~al\mbox{.}(2016)]%
        {Cavallo:2016}
\bibfield{author}{\bibinfo{person}{Andrea Cavallo}, \bibinfo{person}{Atesh Koul}, \bibinfo{person}{Caterina Ansuini}, \bibinfo{person}{Francesca Capozzi}, {and} \bibinfo{person}{Cristina Becchio}.} \bibinfo{year}{2016}\natexlab{}.
\newblock \showarticletitle{Decoding intentions from movement kinematics}.
\newblock \bibinfo{journal}{\emph{Scientific Reports}}  \bibinfo{volume}{6} (\bibinfo{year}{2016}), \bibinfo{pages}{37036 EP --}.
\newblock
\urldef\tempurl%
\url{https://doi.org/10.1038/srep37036}
\showDOI{\tempurl}


\bibitem[Chan et~al\mbox{.}(2015)]%
        {Chan:2015}
\bibfield{author}{\bibinfo{person}{Liwei Chan}, \bibinfo{person}{Yi-Ling Chen}, \bibinfo{person}{Chi-Hao Hsieh}, \bibinfo{person}{Rong-Hao Liang}, {and} \bibinfo{person}{Bing-Yu Chen}.} \bibinfo{year}{2015}\natexlab{}.
\newblock \showarticletitle{CyclopsRing: Enabling Whole-Hand and Context-Aware Interactions Through a Fisheye Ring}. In \bibinfo{booktitle}{\emph{Proceedings of the 28th Annual ACM Symposium on User Interface Software and Technology}} (Charlotte, NC, USA) \emph{(\bibinfo{series}{UIST '15})}. \bibinfo{publisher}{Association for Computing Machinery}, \bibinfo{address}{New York, NY, USA}, \bibinfo{pages}{549–556}.
\newblock
\showISBNx{9781450337793}
\urldef\tempurl%
\url{https://doi.org/10.1145/2807442.2807450}
\showDOI{\tempurl}


\bibitem[Chan et~al\mbox{.}(2013)]%
        {Chan:2013}
\bibfield{author}{\bibinfo{person}{Liwei Chan}, \bibinfo{person}{Rong-Hao Liang}, \bibinfo{person}{Ming-Chang Tsai}, \bibinfo{person}{Kai-Yin Cheng}, \bibinfo{person}{Chao-Huai Su}, \bibinfo{person}{Mike~Y. Chen}, \bibinfo{person}{Wen-Huang Cheng}, {and} \bibinfo{person}{Bing-Yu Chen}.} \bibinfo{year}{2013}\natexlab{}.
\newblock \showarticletitle{FingerPad: Private and Subtle Interaction Using Fingertips}. In \bibinfo{booktitle}{\emph{Proceedings of the 26th Annual ACM Symposium on User Interface Software and Technology}} (St. Andrews, Scotland, United Kingdom) \emph{(\bibinfo{series}{UIST '13})}. \bibinfo{publisher}{Association for Computing Machinery}, \bibinfo{address}{New York, NY, USA}, \bibinfo{pages}{255–260}.
\newblock
\showISBNx{9781450322683}
\urldef\tempurl%
\url{https://doi.org/10.1145/2501988.2502016}
\showDOI{\tempurl}


\bibitem[Chen et~al\mbox{.}(2016)]%
        {Chen:2016}
\bibfield{author}{\bibinfo{person}{Ke-Yu Chen}, \bibinfo{person}{Shwetak~N. Patel}, {and} \bibinfo{person}{Sean Keller}.} \bibinfo{year}{2016}\natexlab{}.
\newblock \showarticletitle{Finexus: Tracking Precise Motions of Multiple Fingertips Using Magnetic Sensing}. In \bibinfo{booktitle}{\emph{Proceedings of the 2016 CHI Conference on Human Factors in Computing Systems}} (San Jose, California, USA) \emph{(\bibinfo{series}{CHI '16})}. \bibinfo{publisher}{Association for Computing Machinery}, \bibinfo{address}{New York, NY, USA}, \bibinfo{pages}{1504–1514}.
\newblock
\showISBNx{9781450333627}
\urldef\tempurl%
\url{https://doi.org/10.1145/2858036.2858125}
\showDOI{\tempurl}


\bibitem[Chen~Chen et~al\mbox{.}(2011)]%
        {Chen:2011}
\bibfield{author}{\bibinfo{person}{F. Chen~Chen}, \bibinfo{person}{A. Favetto}, \bibinfo{person}{M. Mousavi}, \bibinfo{person}{E.~P. Ambrosio}, \bibinfo{person}{S. Appendino}, \bibinfo{person}{Battezzato A.}, \bibinfo{person}{D. Manfredi}, \bibinfo{person}{F. Pescarmona}, {and} \bibinfo{person}{B. Bona}.} \bibinfo{year}{2011}\natexlab{}.
\newblock \showarticletitle{Human Hand: Kinematics, Statics and Dynamics}. In \bibinfo{booktitle}{\emph{International Conference on Environmental Systems}}. \bibinfo{publisher}{AIAA}, \bibinfo{address}{Portland, Oregon}, \bibinfo{pages}{1--10}.
\newblock
\urldef\tempurl%
\url{https://iris.polito.it/handle/11583/2460637}
\showURL{%
\tempurl}


\bibitem[{Cheng} et~al\mbox{.}(2016)]%
        {Cheng:2016}
\bibfield{author}{\bibinfo{person}{H. {Cheng}}, \bibinfo{person}{L. {Yang}}, {and} \bibinfo{person}{Z. {Liu}}.} \bibinfo{year}{2016}\natexlab{}.
\newblock \showarticletitle{Survey on 3D Hand Gesture Recognition}.
\newblock \bibinfo{journal}{\emph{IEEE Transactions on Circuits and Systems for Video Technology}} \bibinfo{volume}{26}, \bibinfo{number}{9} (\bibinfo{date}{Sep.} \bibinfo{year}{2016}), \bibinfo{pages}{1659--1673}.
\newblock
\showISSN{1051-8215}
\urldef\tempurl%
\url{https://doi.org/10.1109/TCSVT.2015.2469551}
\showDOI{\tempurl}


\bibitem[Cheng et~al\mbox{.}(2017)]%
        {unscr14}
\bibfield{author}{\bibinfo{person}{Lung-Pan Cheng}, \bibinfo{person}{Eyal Ofek}, \bibinfo{person}{Christian Holz}, \bibinfo{person}{Hrvoje Benko}, {and} \bibinfo{person}{Andrew~D. Wilson}.} \bibinfo{year}{2017}\natexlab{}.
\newblock \showarticletitle{Sparse Haptic Proxy: Touch Feedback in Virtual Environments Using a General Passive Prop}. In \bibinfo{booktitle}{\emph{Proceedings of the 2017 CHI Conference on Human Factors in Computing Systems}} (Denver, Colorado, USA) \emph{(\bibinfo{series}{CHI '17})}. \bibinfo{publisher}{Association for Computing Machinery}, \bibinfo{address}{New York, NY, USA}, \bibinfo{pages}{3718–3728}.
\newblock
\showISBNx{9781450346559}
\urldef\tempurl%
\url{https://doi.org/10.1145/3025453.3025753}
\showDOI{\tempurl}


\bibitem[Chieffi and Gentilucci(1993)]%
        {Chieffi:1993}
\bibfield{author}{\bibinfo{person}{Sergio Chieffi} {and} \bibinfo{person}{Maurizio Gentilucci}.} \bibinfo{year}{1993}\natexlab{}.
\newblock \showarticletitle{Coordination between the Transport and the Grasp Components During Prehension Movements}.
\newblock \bibinfo{journal}{\emph{Experimental brain research. Experimentelle Hirnforschung. Expérimentation cérébrale}}  \bibinfo{volume}{94} (\bibinfo{date}{02} \bibinfo{year}{1993}), \bibinfo{pages}{471--7}.
\newblock
\urldef\tempurl%
\url{https://doi.org/10.1007/BF00230205}
\showDOI{\tempurl}


\bibitem[Clarence et~al\mbox{.}(2021)]%
        {unscripted}
\bibfield{author}{\bibinfo{person}{Aldrich Clarence}, \bibinfo{person}{Jarrod Knibbe}, \bibinfo{person}{Maxime Cordeil}, {and} \bibinfo{person}{Michael Wybrow}.} \bibinfo{year}{2021}\natexlab{}.
\newblock \showarticletitle{Unscripted Retargeting: Reach Prediction for Haptic Retargeting in Virtual Reality}. In \bibinfo{booktitle}{\emph{2021 IEEE Virtual Reality and 3D User Interfaces (VR)}}. \bibinfo{pages}{150--159}.
\newblock
\urldef\tempurl%
\url{https://doi.org/10.1109/VR50410.2021.00036}
\showDOI{\tempurl}


\bibitem[Clarence et~al\mbox{.}(2022)]%
        {Clarence:2022}
\bibfield{author}{\bibinfo{person}{Aldrich Clarence}, \bibinfo{person}{Jarrod Knibbe}, \bibinfo{person}{Maxime Cordeil}, {and} \bibinfo{person}{Michael Wybrow}.} \bibinfo{year}{2022}\natexlab{}.
\newblock \showarticletitle{Investigating The Effect of Direction on The Limits of Haptic Retargeting}. In \bibinfo{booktitle}{\emph{2022 IEEE International Symposium on Mixed and Augmented Reality (ISMAR)}}. \bibinfo{pages}{612--621}.
\newblock
\urldef\tempurl%
\url{https://doi.org/10.1109/ISMAR55827.2022.00078}
\showDOI{\tempurl}


\bibitem[Daiber et~al\mbox{.}(2012)]%
        {Daiber2012}
\bibfield{author}{\bibinfo{person}{Florian Daiber}, \bibinfo{person}{Dimitar Valkov}, \bibinfo{person}{Frank Steinicke}, \bibinfo{person}{Antonio Kr\"uger}, {and} \bibinfo{person}{Klaus~H. Hinrichs}.} \bibinfo{year}{2012}\natexlab{}.
\newblock \showarticletitle{Towards Object Prediction based on Hand Postures for Reach to Grasp Interaction}. In \bibinfo{booktitle}{\emph{Proceedings of the ACM CHI 2012 Workshop on Touching the 3rd Dimension of CHI: Touching and Designing 3D User Interfaces (3DCHI)}}. \bibinfo{pages}{99--106}.
\newblock


\bibitem[Della~Santina et~al\mbox{.}(2017)]%
        {DellaSantina:2017}
\bibfield{author}{\bibinfo{person}{Cosimo Della~Santina}, \bibinfo{person}{Matteo Bianchi}, \bibinfo{person}{Giuseppe Averta}, \bibinfo{person}{Simone Ciotti}, \bibinfo{person}{Visar Arapi}, \bibinfo{person}{Simone Fani}, \bibinfo{person}{Edoardo Battaglia}, \bibinfo{person}{Manuel Giuseppe~Catalano}, \bibinfo{person}{Marco Santello}, {and} \bibinfo{person}{Antonio Bicchi}.} \bibinfo{year}{2017}\natexlab{}.
\newblock \showarticletitle{Postural Hand Synergies during Environmental Constraint Exploitation}.
\newblock \bibinfo{journal}{\emph{Frontiers in Neurorobotics}}  \bibinfo{volume}{11} (\bibinfo{date}{08} \bibinfo{year}{2017}).
\newblock
\urldef\tempurl%
\url{https://doi.org/10.3389/fnbot.2017.00041}
\showDOI{\tempurl}


\bibitem[Duncan et~al\mbox{.}(2013)]%
        {Duncan:2013}
\bibfield{author}{\bibinfo{person}{Scott~F.M. Duncan}, \bibinfo{person}{Caitlin~E. Saracevic}, {and} \bibinfo{person}{Ryosuke Kakinoki}.} \bibinfo{year}{2013}\natexlab{}.
\newblock \showarticletitle{Biomechanics of the Hand}.
\newblock \bibinfo{journal}{\emph{Hand Clinics}} \bibinfo{volume}{29}, \bibinfo{number}{4} (\bibinfo{date}{nov} \bibinfo{year}{2013}), \bibinfo{pages}{483--492}.
\newblock
\urldef\tempurl%
\url{https://doi.org/10.1016/j.hcl.2013.08.003}
\showDOI{\tempurl}


\bibitem[Egmose and Køppe(2018)]%
        {Egmose:2018}
\bibfield{author}{\bibinfo{person}{Ida Egmose} {and} \bibinfo{person}{Simo Køppe}.} \bibinfo{year}{2018}\natexlab{}.
\newblock \showarticletitle{Shaping of Reach-to-Grasp Kinematics by Intentions: A Meta-Analysis}.
\newblock \bibinfo{journal}{\emph{Journal of Motor Behavior}} \bibinfo{volume}{50}, \bibinfo{number}{2} (\bibinfo{year}{2018}), \bibinfo{pages}{155--165}.
\newblock
\urldef\tempurl%
\url{https://doi.org/10.1080/00222895.2017.1327407}
\showDOI{\tempurl}
\showeprint{https://doi.org/10.1080/00222895.2017.1327407}
\newblock
\shownote{PMID: 28644719}.


\bibitem[El-Khoury et~al\mbox{.}(2013)]%
        {robot-grasp-model}
\bibfield{author}{\bibinfo{person}{Sahar El-Khoury}, \bibinfo{person}{Miao Li}, {and} \bibinfo{person}{Aude Billard}.} \bibinfo{year}{2013}\natexlab{}.
\newblock \showarticletitle{On the generation of a variety of grasps}.
\newblock \bibinfo{journal}{\emph{Robotics and Autonomous Systems}} \bibinfo{volume}{61}, \bibinfo{number}{12} (\bibinfo{year}{2013}), \bibinfo{pages}{1335--1349}.
\newblock
\showISSN{0921-8890}
\urldef\tempurl%
\url{https://doi.org/10.1016/j.robot.2013.08.002}
\showDOI{\tempurl}


\bibitem[Elliott(1987)]%
        {elliott1987handbook}
\bibfield{author}{\bibinfo{person}{D.F. Elliott}.} \bibinfo{year}{1987}\natexlab{}.
\newblock \bibinfo{booktitle}{\emph{Handbook of Digital Signal Processing: Engineering Applications}}.
\newblock \bibinfo{publisher}{Academic Press}.
\newblock
\showISBNx{9780122370755}
\showLCCN{86026490}
\urldef\tempurl%
\url{https://books.google.de/books?id=7s5yQgAACAAJ}
\showURL{%
\tempurl}


\bibitem[Feix et~al\mbox{.}(2014a)]%
        {Feix2014grasptask}
\bibfield{author}{\bibinfo{person}{Thomas Feix}, \bibinfo{person}{Ian~M. Bullock}, {and} \bibinfo{person}{Aaron~M. Dollar}.} \bibinfo{year}{2014}\natexlab{a}.
\newblock \showarticletitle{Analysis of Human Grasping Behavior: Correlating Tasks, Objects and Grasps}.
\newblock \bibinfo{journal}{\emph{IEEE Transactions on Haptics}} \bibinfo{volume}{7}, \bibinfo{number}{4} (\bibinfo{date}{Oct} \bibinfo{year}{2014}), \bibinfo{pages}{430--441}.
\newblock
\showISSN{2329-4051}
\urldef\tempurl%
\url{https://doi.org/10.1109/TOH.2014.2326867}
\showDOI{\tempurl}


\bibitem[Feix et~al\mbox{.}(2014b)]%
        {Feix2014graspobject}
\bibfield{author}{\bibinfo{person}{Thomas Feix}, \bibinfo{person}{Ian~M. Bullock}, {and} \bibinfo{person}{Aaron~M. Dollar}.} \bibinfo{year}{2014}\natexlab{b}.
\newblock \showarticletitle{Analysis of Human Grasping Behavior: Object Characteristics and Grasp Type}.
\newblock \bibinfo{journal}{\emph{IEEE Transactions on Haptics}} \bibinfo{volume}{7}, \bibinfo{number}{3} (\bibinfo{date}{July} \bibinfo{year}{2014}), \bibinfo{pages}{311--323}.
\newblock
\showISSN{2329-4051}
\urldef\tempurl%
\url{https://doi.org/10.1109/TOH.2014.2326871}
\showDOI{\tempurl}


\bibitem[{Fu} and {Santello}(2011)]%
        {Fu:2011}
\bibfield{author}{\bibinfo{person}{Q. {Fu}} {and} \bibinfo{person}{M. {Santello}}.} \bibinfo{year}{2011}\natexlab{}.
\newblock \showarticletitle{Towards a complete description of grasping kinematics: A framework for quantifying human grasping and manipulation}. In \bibinfo{booktitle}{\emph{2011 Annual International Conference of the IEEE Engineering in Medicine and Biology Society}}. \bibinfo{pages}{8247--8250}.
\newblock
\showISSN{1558-4615}
\urldef\tempurl%
\url{https://doi.org/10.1109/IEMBS.2011.6092033}
\showDOI{\tempurl}


\bibitem[Furmanek et~al\mbox{.}(2019)]%
        {Furmanek:2019}
\bibfield{author}{\bibinfo{person}{M.P. Furmanek}, \bibinfo{person}{L.F. Schettino}, {and} \bibinfo{person}{M. Yarossi}.} \bibinfo{year}{2019}\natexlab{}.
\newblock \showarticletitle{{Coordination of reach-to-grasp in physical and haptic-free virtual environments}}.
\newblock \bibinfo{journal}{\emph{Journal of NeuroEngineering Rehabilitation}} \bibinfo{volume}{16}, \bibinfo{number}{78} (\bibinfo{year}{2019}).
\newblock
\urldef\tempurl%
\url{https://doi.org/10.1186/s12984-019-0525-9}
\showDOI{\tempurl}


\bibitem[Gonzalez and Follmer(2023)]%
        {recent-CHI}
\bibfield{author}{\bibinfo{person}{Eric~J Gonzalez} {and} \bibinfo{person}{Sean Follmer}.} \bibinfo{year}{2023}\natexlab{}.
\newblock \showarticletitle{Sensorimotor Simulation of Redirected Reaching Using Stochastic Optimal Feedback Control}. In \bibinfo{booktitle}{\emph{Proceedings of the 2023 CHI Conference on Human Factors in Computing Systems}} (Hamburg, Germany) \emph{(\bibinfo{series}{CHI '23})}. \bibinfo{publisher}{Association for Computing Machinery}, \bibinfo{address}{New York, NY, USA}, Article \bibinfo{articleno}{776}, \bibinfo{numpages}{17}~pages.
\newblock
\showISBNx{9781450394215}
\urldef\tempurl%
\url{https://doi.org/10.1145/3544548.3580767}
\showDOI{\tempurl}


\bibitem[Henrikson et~al\mbox{.}(2020)]%
        {unscr23}
\bibfield{author}{\bibinfo{person}{Rorik Henrikson}, \bibinfo{person}{Tovi Grossman}, \bibinfo{person}{Sean Trowbridge}, \bibinfo{person}{Daniel Wigdor}, {and} \bibinfo{person}{Hrvoje Benko}.} \bibinfo{year}{2020}\natexlab{}.
\newblock \showarticletitle{Head-Coupled Kinematic Template Matching: A Prediction Model for Ray Pointing in VR}. In \bibinfo{booktitle}{\emph{Proceedings of the 2020 CHI Conference on Human Factors in Computing Systems}} (Honolulu, HI, USA) \emph{(\bibinfo{series}{CHI '20})}. \bibinfo{publisher}{Association for Computing Machinery}, \bibinfo{address}{New York, NY, USA}, \bibinfo{pages}{1–14}.
\newblock
\showISBNx{9781450367080}
\urldef\tempurl%
\url{https://doi.org/10.1145/3313831.3376489}
\showDOI{\tempurl}


\bibitem[Henze et~al\mbox{.}(2017)]%
        {unscr25}
\bibfield{author}{\bibinfo{person}{Niels Henze}, \bibinfo{person}{Sven Mayer}, \bibinfo{person}{Huy~Viet Le}, {and} \bibinfo{person}{Valentin Schwind}.} \bibinfo{year}{2017}\natexlab{}.
\newblock \showarticletitle{Improving Software-Reduced Touchscreen Latency}. In \bibinfo{booktitle}{\emph{Proceedings of the 19th International Conference on Human-Computer Interaction with Mobile Devices and Services}} (Vienna, Austria) \emph{(\bibinfo{series}{MobileHCI '17})}. \bibinfo{publisher}{Association for Computing Machinery}, \bibinfo{address}{New York, NY, USA}, Article \bibinfo{articleno}{107}, \bibinfo{numpages}{8}~pages.
\newblock
\showISBNx{9781450350754}
\urldef\tempurl%
\url{https://doi.org/10.1145/3098279.3122150}
\showDOI{\tempurl}


\bibitem[Heumer et~al\mbox{.}(2008)]%
        {Heumer:2008}
\bibfield{author}{\bibinfo{person}{Guido Heumer}, \bibinfo{person}{Heni~Ben Amor}, {and} \bibinfo{person}{Bernhard Jung}.} \bibinfo{year}{2008}\natexlab{}.
\newblock \showarticletitle{Grasp Recognition for Uncalibrated Data Gloves: A Machine Learning Approach}.
\newblock \bibinfo{journal}{\emph{Presence}} \bibinfo{volume}{17}, \bibinfo{number}{2} (\bibinfo{year}{2008}), \bibinfo{pages}{121--142}.
\newblock
\urldef\tempurl%
\url{https://doi.org/10.1162/pres.17.2.121}
\showDOI{\tempurl}


\bibitem[{Höll} et~al\mbox{.}(2018)]%
        {Hoell:2018}
\bibfield{author}{\bibinfo{person}{M. {Höll}}, \bibinfo{person}{M. {Oberweger}}, \bibinfo{person}{C. {Arth}}, {and} \bibinfo{person}{V. {Lepetit}}.} \bibinfo{year}{2018}\natexlab{}.
\newblock \showarticletitle{Efficient Physics-Based Implementation for Realistic Hand-Object Interaction in Virtual Reality}. In \bibinfo{booktitle}{\emph{2018 IEEE Conference on Virtual Reality and 3D User Interfaces (VR)}}. \bibinfo{pages}{175--182}.
\newblock
\urldef\tempurl%
\url{https://doi.org/10.1109/VR.2018.8448284}
\showDOI{\tempurl}


\bibitem[Kang and Ikeuchi(1994)]%
        {Sing:1994}
\bibfield{author}{\bibinfo{person}{Sing~Bing Kang} {and} \bibinfo{person}{K. Ikeuchi}.} \bibinfo{year}{1994}\natexlab{}.
\newblock \showarticletitle{Determination of motion breakpoints in a task sequence from human hand motion}. In \bibinfo{booktitle}{\emph{Proceedings of the 1994 IEEE International Conference on Robotics and Automation}}. \bibinfo{pages}{551--556 vol.1}.
\newblock
\urldef\tempurl%
\url{https://doi.org/10.1109/ROBOT.1994.351241}
\showDOI{\tempurl}


\bibitem[Lank et~al\mbox{.}(2007)]%
        {unscr30}
\bibfield{author}{\bibinfo{person}{Edward Lank}, \bibinfo{person}{Yi-Chun~Nikko Cheng}, {and} \bibinfo{person}{Jaime Ruiz}.} \bibinfo{year}{2007}\natexlab{}.
\newblock \showarticletitle{Endpoint Prediction Using Motion Kinematics}. In \bibinfo{booktitle}{\emph{Proceedings of the SIGCHI Conference on Human Factors in Computing Systems}} (San Jose, California, USA) \emph{(\bibinfo{series}{CHI '07})}. \bibinfo{publisher}{Association for Computing Machinery}, \bibinfo{address}{New York, NY, USA}, \bibinfo{pages}{637–646}.
\newblock
\showISBNx{9781595935939}
\urldef\tempurl%
\url{https://doi.org/10.1145/1240624.1240724}
\showDOI{\tempurl}


\bibitem[MacKenzie and Iberall(1994)]%
        {MacKenzie:1994}
\bibfield{author}{\bibinfo{person}{Christine~L MacKenzie} {and} \bibinfo{person}{Thea Iberall}.} \bibinfo{year}{1994}\natexlab{}.
\newblock \bibinfo{booktitle}{\emph{The grasping hand}}.
\newblock \bibinfo{publisher}{Amsterdam ; New York : North-Holland}.
\newblock
\showISBNx{0444817468}


\bibitem[Marwecki et~al\mbox{.}(2019)]%
        {unscr35}
\bibfield{author}{\bibinfo{person}{Sebastian Marwecki}, \bibinfo{person}{Andrew~D. Wilson}, \bibinfo{person}{Eyal Ofek}, \bibinfo{person}{Mar Gonzalez~Franco}, {and} \bibinfo{person}{Christian Holz}.} \bibinfo{year}{2019}\natexlab{}.
\newblock \showarticletitle{Mise-Unseen: Using Eye Tracking to Hide Virtual Reality Scene Changes in Plain Sight}. In \bibinfo{booktitle}{\emph{Proceedings of the 32nd Annual ACM Symposium on User Interface Software and Technology}} (New Orleans, LA, USA) \emph{(\bibinfo{series}{UIST '19})}. \bibinfo{publisher}{Association for Computing Machinery}, \bibinfo{address}{New York, NY, USA}, \bibinfo{pages}{777–789}.
\newblock
\showISBNx{9781450368162}
\urldef\tempurl%
\url{https://doi.org/10.1145/3332165.3347919}
\showDOI{\tempurl}


\bibitem[Molina-Vilaplana et~al\mbox{.}(2002)]%
        {Molina-Vilaplana:2002}
\bibfield{author}{\bibinfo{person}{Javier Molina-Vilaplana}, \bibinfo{person}{Jorge~Feliu Batlle}, {and} \bibinfo{person}{Juan~L{\'o}pez Coronado}.} \bibinfo{year}{2002}\natexlab{}.
\newblock \showarticletitle{A Neural Model of Spatio Temporal Coordination in Prehension}. In \bibinfo{booktitle}{\emph{Artificial Neural Networks --- ICANN 2002}}, \bibfield{editor}{\bibinfo{person}{Jos{\'e}~R. Dorronsoro}} (Ed.). \bibinfo{publisher}{Springer Berlin Heidelberg}, \bibinfo{address}{Berlin, Heidelberg}, \bibinfo{pages}{9--14}.
\newblock
\showISBNx{978-3-540-46084-8}


\bibitem[Pasqual and Wobbrock(2014)]%
        {unscr40}
\bibfield{author}{\bibinfo{person}{Phillip~T. Pasqual} {and} \bibinfo{person}{Jacob~O. Wobbrock}.} \bibinfo{year}{2014}\natexlab{}.
\newblock \showarticletitle{Mouse Pointing Endpoint Prediction Using Kinematic Template Matching}. In \bibinfo{booktitle}{\emph{Proceedings of the SIGCHI Conference on Human Factors in Computing Systems}} (Toronto, Ontario, Canada) \emph{(\bibinfo{series}{CHI '14})}. \bibinfo{publisher}{Association for Computing Machinery}, \bibinfo{address}{New York, NY, USA}, \bibinfo{pages}{743–752}.
\newblock
\showISBNx{9781450324731}
\urldef\tempurl%
\url{https://doi.org/10.1145/2556288.2557406}
\showDOI{\tempurl}


\bibitem[Paulson et~al\mbox{.}(2011)]%
        {Paulson:2011}
\bibfield{author}{\bibinfo{person}{B. Paulson}, \bibinfo{person}{D. Cummings}, {and} \bibinfo{person}{T. Hammond}.} \bibinfo{year}{2011}\natexlab{}.
\newblock \showarticletitle{Object interaction detection using hand posture cues in an office setting}.
\newblock \bibinfo{journal}{\emph{International Journal of Human Computer Studies}} \bibinfo{volume}{69}, \bibinfo{number}{1-2} (\bibinfo{year}{2011}), \bibinfo{pages}{19--29}.
\newblock
\showCODEN{IHSTE}
\showISSN{10715819}
\urldef\tempurl%
\url{https://doi.org/10.1016/j.ijhcs.2010.09.003}
\showDOI{\tempurl}


\bibitem[Pham et~al\mbox{.}(2017)]%
        {PHAM2017218}
\bibfield{author}{\bibinfo{person}{Trang Pham}, \bibinfo{person}{Truyen Tran}, \bibinfo{person}{Dinh Phung}, {and} \bibinfo{person}{Svetha Venkatesh}.} \bibinfo{year}{2017}\natexlab{}.
\newblock \showarticletitle{Predicting healthcare trajectories from medical records: A deep learning approach}.
\newblock \bibinfo{journal}{\emph{Journal of Biomedical Informatics}}  \bibinfo{volume}{69} (\bibinfo{year}{2017}), \bibinfo{pages}{218--229}.
\newblock
\showISSN{1532-0464}
\urldef\tempurl%
\url{https://doi.org/10.1016/j.jbi.2017.04.001}
\showDOI{\tempurl}


\bibitem[Sak et~al\mbox{.}(2014)]%
        {DBLP:journals/corr/SakSB14}
\bibfield{author}{\bibinfo{person}{Hasim Sak}, \bibinfo{person}{Andrew~W. Senior}, {and} \bibinfo{person}{Fran{\c{c}}oise Beaufays}.} \bibinfo{year}{2014}\natexlab{}.
\newblock \showarticletitle{Long Short-Term Memory Based Recurrent Neural Network Architectures for Large Vocabulary Speech Recognition}.
\newblock \bibinfo{journal}{\emph{CoRR}}  \bibinfo{volume}{abs/1402.1128} (\bibinfo{year}{2014}), \bibinfo{pages}{1--10}.
\newblock
\showeprint[arXiv]{1402.1128}
\urldef\tempurl%
\url{http://arxiv.org/abs/1402.1128}
\showURL{%
\tempurl}


\bibitem[Santello et~al\mbox{.}(2016)]%
        {Santello:2016}
\bibfield{author}{\bibinfo{person}{Marco Santello}, \bibinfo{person}{Matteo Bianchi}, \bibinfo{person}{Marco Gabiccini}, \bibinfo{person}{Emiliano Ricciardi}, \bibinfo{person}{Gionata Salvietti}, \bibinfo{person}{Domenico Prattichizzo}, \bibinfo{person}{Marc Ernst}, \bibinfo{person}{Alessandro Moscatelli}, \bibinfo{person}{Henrik Jörntell}, \bibinfo{person}{Astrid~M.L. Kappers}, \bibinfo{person}{Kostas Kyriakopoulos}, \bibinfo{person}{Alin Albu-Schäffer}, \bibinfo{person}{Claudio Castellini}, {and} \bibinfo{person}{Antonio Bicchi}.} \bibinfo{year}{2016}\natexlab{}.
\newblock \showarticletitle{Hand synergies: Integration of robotics and neuroscience for understanding the control of biological and artificial hands}.
\newblock \bibinfo{journal}{\emph{Physics of Life Reviews}}  \bibinfo{volume}{17} (\bibinfo{year}{2016}), \bibinfo{pages}{1 -- 23}.
\newblock
\showISSN{1571-0645}
\urldef\tempurl%
\url{https://doi.org/10.1016/j.plrev.2016.02.001}
\showDOI{\tempurl}


\bibitem[Santello and Soechting(1998)]%
        {Santello:1998}
\bibfield{author}{\bibinfo{person}{M. Santello} {and} \bibinfo{person}{J.~F. Soechting}.} \bibinfo{year}{1998}\natexlab{}.
\newblock \showarticletitle{Gradual molding of the hand to object contours}.
\newblock \bibinfo{journal}{\emph{Journal of neurophysiology}} \bibinfo{volume}{79}, \bibinfo{number}{3} (\bibinfo{year}{1998}), \bibinfo{pages}{1307--1320}.
\newblock
\showISSN{0022-3077}
\urldef\tempurl%
\url{https://doi.org/10.1152/jn.1998.79.3.1307}
\showDOI{\tempurl}


\bibitem[Taylor et~al\mbox{.}(2016)]%
        {Taylor:2016}
\bibfield{author}{\bibinfo{person}{Jonathan Taylor}, \bibinfo{person}{Lucas Bordeaux}, \bibinfo{person}{Thomas Cashman}, \bibinfo{person}{Bob Corish}, \bibinfo{person}{Cem Keskin}, \bibinfo{person}{Toby Sharp}, \bibinfo{person}{Eduardo Soto}, \bibinfo{person}{David Sweeney}, \bibinfo{person}{Julien Valentin}, \bibinfo{person}{Benjamin Luff}, \bibinfo{person}{Arran Topalian}, \bibinfo{person}{Erroll Wood}, \bibinfo{person}{Sameh Khamis}, \bibinfo{person}{Pushmeet Kohli}, \bibinfo{person}{Shahram Izadi}, \bibinfo{person}{Richard Banks}, \bibinfo{person}{Andrew Fitzgibbon}, {and} \bibinfo{person}{Jamie Shotton}.} \bibinfo{year}{2016}\natexlab{}.
\newblock \showarticletitle{Efficient and Precise Interactive Hand Tracking Through Joint, Continuous Optimization of Pose and Correspondences}.
\newblock \bibinfo{journal}{\emph{ACM Trans. Graph.}} \bibinfo{volume}{35}, \bibinfo{number}{4}, Article \bibinfo{articleno}{143} (\bibinfo{date}{July} \bibinfo{year}{2016}), \bibinfo{numpages}{12}~pages.
\newblock
\showISSN{0730-0301}
\urldef\tempurl%
\url{https://doi.org/10.1145/2897824.2925965}
\showDOI{\tempurl}


\bibitem[Vatavu and Zaii(2013)]%
        {Vatavu2013a}
\bibfield{author}{\bibinfo{person}{Radu-Daniel Vatavu} {and} \bibinfo{person}{Ionu~Alexandru Zaii}.} \bibinfo{year}{2013}\natexlab{}.
\newblock \showarticletitle{Automatic Recognition of Object Size and Shape via User-dependent Measurements of the Grasping Hand}.
\newblock \bibinfo{journal}{\emph{Int. J. Hum.-Comput. Stud.}} \bibinfo{volume}{71}, \bibinfo{number}{5} (\bibinfo{date}{May} \bibinfo{year}{2013}), \bibinfo{pages}{590--607}.
\newblock
\showISSN{1071-5819}
\urldef\tempurl%
\url{https://doi.org/10.1016/j.ijhcs.2013.01.002}
\showDOI{\tempurl}


\bibitem[Xia et~al\mbox{.}(2014)]%
        {Xia:2014}
\bibfield{author}{\bibinfo{person}{Haijun Xia}, \bibinfo{person}{Ricardo Jota}, \bibinfo{person}{Benjamin McCanny}, \bibinfo{person}{Zhe Yu}, \bibinfo{person}{Clifton Forlines}, \bibinfo{person}{Karan Singh}, {and} \bibinfo{person}{Daniel Wigdor}.} \bibinfo{year}{2014}\natexlab{}.
\newblock \showarticletitle{Zero-Latency Tapping: Using Hover Information to Predict Touch Locations and Eliminate Touchdown Latency}. In \bibinfo{booktitle}{\emph{Proceedings of the 27th Annual ACM Symposium on User Interface Software and Technology}} (Honolulu, Hawaii, USA) \emph{(\bibinfo{series}{UIST '14})}. \bibinfo{publisher}{Association for Computing Machinery}, \bibinfo{address}{New York, NY, USA}, \bibinfo{pages}{205–214}.
\newblock
\showISBNx{9781450330695}
\urldef\tempurl%
\url{https://doi.org/10.1145/2642918.2647348}
\showDOI{\tempurl}


\bibitem[Yu et~al\mbox{.}(2019)]%
        {unscr48}
\bibfield{author}{\bibinfo{person}{Difeng Yu}, \bibinfo{person}{Hai-Ning Liang}, \bibinfo{person}{Xueshi Lu}, \bibinfo{person}{Kaixuan Fan}, {and} \bibinfo{person}{Barrett Ens}.} \bibinfo{year}{2019}\natexlab{}.
\newblock \showarticletitle{Modeling endpoint distribution of pointing selection tasks in virtual reality environments}.
\newblock \bibinfo{journal}{\emph{ACM Transactions on Graphics (TOG)}} \bibinfo{volume}{38}, \bibinfo{number}{6} (\bibinfo{year}{2019}), \bibinfo{pages}{1--13}.
\newblock


\end{thebibliography}

\end{document}